\documentclass[aps,twocolumn,floatfix]{revtex4-2}
\usepackage[utf8]{inputenc}

\usepackage{amsmath,amssymb}
\usepackage{color}
\usepackage{graphicx}
\usepackage{tikz}
\usepackage{pgfplots}
\usepackage{hyperref}

\begin{document}
\title{Holographic dark energy in the generalized mass-to-horizon cosmology}

\author{H. Aghanoor Ashtiani\textsuperscript{1}}
\author{A. Sheykhi\textsuperscript{1,2}}
\email{asheykhi@shirazu.ac.ir}
\author{Amir A.~Khodahami \textsuperscript{1,2}}
\email{a.khodahami@shirazu.ac.ir}
\affiliation{\textsuperscript{1}Department of Physics, College of Science, Shiraz University, Shiraz 71454, Iran}
\affiliation{\textsuperscript{2}Biruni Observatory, College of Science, Shiraz University, Shiraz 71454, Iran}

\begin{abstract}
We investigate a holographic dark energy (HDE) model based on
generalized mass-to-horizon entropy (GMHE), incorporating the
entropy corrections simultaneously into the Friedmann equation and
the dark-energy density. Two infrared (IR) cutoffs are examined: the Hubble radius and the Granda--Oliveros (GO) scale. For the Hubble-radius
cutoff, non-interacting HDE behaves as pressureless matter and
cannot drive acceleration, whereas a dark-sector interaction
allows accelerated expansion. For the GO cutoff, the analysis
covers dark-energy-dominated, non-interacting, and interacting
regimes. We derive analytical solutions in the dark-energy-dominated limit and examine the evolution of the dark-energy density parameter, equation of state, and deceleration parameter in the two-component scenarios. For suitable parameter
choices, the GO model reproduces a transition from decelerated to
accelerated expansion at $z\approx 0.6$. The squared-sound-speed analysis shows that the stability results are sensitive to the GO cutoff parameters and interaction strength; divergences occur for some parameter choices. The statefinder
diagnostic further reveals that departures from $\Lambda$CDM
depend on the entropy modification, the GO cutoff parameters, and
the dark-sector interaction. Some trajectories approach the
$\Lambda$CDM reference point at the present epoch while retaining
distinct behavior at earlier times. These results show that the GO cutoff produces a wider range of background evolution in the GMHE framework, motivating further tests of its observational viability.
\end{abstract}
 \maketitle
 \newpage
\section{Introduction}
Observations of Type Ia supernovae \cite{Riess1998, Perlmutter1999}, later supported by cosmic microwave background \cite{Planck2020} and large-scale structure data \cite{BOSS2017, DES2024}, established that the universe is undergoing accelerated expansion. In general relativity, this acceleration is commonly attributed to dark energy, which has negative pressure and contributes substantially to the current cosmic energy budget. The simplest dark-energy model is a cosmological constant, $\Lambda$, with $w_{\Lambda}=-1$. Although the $\Lambda$CDM framework fits a broad range of observations, it is associated with the fine-tuning \cite{Weinberg1989} and coincidence \cite{Zlatev1999} problems. These issues have motivated studies of dynamical dark-energy models, in which the density and equation of state evolve with time.

Holographic dark energy (HDE) is a prominent approach to dynamical dark energy \cite{Li2004}. Motivated by the holographic principle \cite{tHooft1993, Susskind1995}, which bounds the degrees of freedom by the boundary area, it relates the energy density associated with a cosmological length scale $L$ to that scale \cite{Cohen1999}. The resulting dark energy
density takes the form
\begin{equation}
\label{eq:rhoDE}
\rho_{\rm DE} = 3 c^2 M_{\rm Pl}^2 L^{-2},
\end{equation}
where $c^2$ is a dimensionless parameter and $L$ is the size or
radius of the horizon under consideration, commonly identified
with the infrared (IR) cutoff. The IR cutoff determines the HDE density and therefore affects the predicted equation of state and expansion history. Common choices include the Hubble radius, future event horizon, and particle horizon, each leading to distinct cosmological dynamics that can be tested observationally \cite{Li2004, AbdollahiZadeh2018}. A related line of work connects gravitational field equations with thermodynamics \cite{Jacobson1995, Padmanabhan2011}. In FRW cosmology, applying the first law of thermodynamics to the apparent horizon reproduces the Friedmann equations \cite{CaiKim2005}. While the standard Bekenstein-Hawking area law $S = A/(4G)$ yields standard general relativity \cite{Bekenstein1973, Hawking1975}, quantum-gravity corrections modify the horizon entropy \cite{Cai2008}, thereby altering the resulting cosmological field equations \cite{Chagoya2025}.

Recently, a generalized mass-to-horizon entropy (GMHE) was proposed based on a modified mass-to-horizon relation \cite{Gohar2024},
    \begin{equation}
        M = \gamma \frac{c^2 L^n}{G}, \label{eq:mass-horizon}
    \end{equation}
where $n$ is a non-negative real parameter and $\gamma$ has dimensions of $\textit{length}^{1-n}$. The GMHE was introduced to resolve a thermodynamic inconsistency that can arise when generalized entropies are combined with the Hawking temperature and the Clausius relation. Subsequent investigations have demonstrated that this framework provides a viable cosmological description of cosmic acceleration and encompasses several known generalized entropy models, including Barrow, Tsallis-Cirto, and Tsallis-Zamora entropies, as special limits \cite{Gohar2026, Sheykhi2026}.

Research on GMHE has developed along three primary directions. First, studies have developed its theoretical foundations, including the generalized mass-to-horizon relation, the Clausius-consistent entropy, and the thermodynamic basis of entropic cosmology \cite{Gohar2024, GoharSalzano2024, Gohar2026, Sheykhi2026, Mondal2026, Shameeem2026, Saridakis2026}. Second, cosmological applications have derived modified Friedmann equations from the apparent-horizon thermodynamics and investigated the background dynamics through the Hubble parameter, equation of state, and deceleration parameter \cite{Sheykhi2026, Shameeem2026, Prasanthan2026a, Basilakos2025}. Third, phenomenological studies have explored observational constraints and cosmological implications, including structure growth, gravitational waves, baryogenesis, and BAO data \cite{Basilakos2025, Luciano2025, Denkiewicz2026, Prasanthan2026b, Luciano2026a, LucianoSaridakis2026, Sheykhi2026b, Mondal2026b}.

Here we study HDE in a GMHE cosmology, applying the entropy modification consistently to both the background Friedmann equation and the HDE density. The resulting density scales as $\rho_D\propto L^{n-3}$, rather than the standard $L^{-2}$ scaling. We investigate this framework using both the Hubble radius and Granda--Oliveros (GO) cutoffs in non-interacting and interacting dark sectors, deriving analytical solutions in the dark-energy-dominated limit for the GO cutoff and assessing the cosmological viability through background, sound-speed, and statefinder analyses.

The rest of this paper is structured as follows: Sec.~\ref{First} reviews the modified Friedmann equations derived from the mass-to-horizon entropy. Sec.~\ref{cosm} introduces the HDE density with the Hubble-radius cutoff and establishes the background dynamical equations. Sec.~\ref{GOasIR} analyzes the cosmological dynamics with Hubble and Granda--Oliveros cutoffs in GMHE cosmology. Sec.~\ref{Stab} presents the stability analysis and statefinder diagnostics. Finally, Sec.~\ref{Con} summarizes our conclusions.
\section{Modified Cosmology through mass-to-horizon entropy\label{First}}
We consider a spatially homogeneous and isotropic FRW universe described by the line element
\begin{equation}
ds^2={h}_{\mu \nu}dx^{\mu}
dx^{\nu}+\tilde{r}^2(d\theta^2+\sin^2\theta d\phi^2),
\end{equation}
where $a(t)$ is the scale factor, $\tilde{r}=a(t)r$. The curvature parameter $k$ takes the values $-1, 0$, and $1$ for open, flat, and closed universes, respectively. Here we take $x^0=t,
x^1=r$, and $h_{\mu \nu}$=diag $(-1, a^2/(1-kr^2))$. Several horizons are defined in cosmology; however, the apparent horizon is the most widely adopted and consistent boundary in thermodynamic treatments of FRW cosmology. In the
background of FRW universe, the radius of the apparent horizon is
determined via $h^{\mu
\nu}\partial_{\mu}\tilde{r}_A\partial_{\nu}\tilde{r}_A=0$, which
implies that the vector $\nabla \tilde{r}_A$ is null on the
apparent horizon surface. The apparent horizon radius is therefore given by \cite{Hay1,Hay2,Bak}
\begin{equation}\label{radius}
 \tilde{r}_A=\frac{1}{\sqrt{H^2+k/a^2}},
\end{equation}
where $H=\dot{a}/a$ is the Hubble parameter. We treat the universe as a thermodynamic system bounded by the apparent horizon. In analogy with black hole thermodynamics, a surface gravity and an associated temperature can be assigned to this horizon. The surface gravity of the apparent horizon is defined as
\begin{equation}\label{kappa}
\kappa = \frac{1}{2\sqrt{-h}} \partial_{\mu} \left( \sqrt{-h}
h^{\mu \nu}
\partial_{\nu} \tilde{r}_A \right).
\end{equation}
The associated temperature with the apparent horizon is obtained
as
\begin{equation}\label{T}
T_h=\frac{\kappa}{2\pi}=-\frac{1}{2 \pi \tilde
r_A}\left(1-\frac{\dot {\tilde r}_A}{2H\tilde r_A}\right).
\end{equation}
We model the cosmic energy content as a perfect fluid with energy-momentum tensor
\begin{equation}\label{TEM}
    T_{\mu\nu}=(\rho+p)u_{\mu}u_{\nu}+pg_{\mu\nu},
\end{equation}
where $\rho$ and $p$ denote the energy density and pressure, respectively. The energy-momentum conservation law, $\nabla_{\mu}T^{\mu\nu}=0$, in FRW spacetime yields the continuity equation
\begin{equation}\label{Cons}
    \dot{\rho}+3H(\rho+p)=0.
\end{equation}
The work density associated with volume change is defined as \cite{Hay2}
\begin{equation}\label{Work}
    W=-\frac{1}{2} T^{\mu\nu}h_{\mu\nu},
\end{equation}
which gives
\begin{equation}\label{Work2}
    W=\frac{1}{2}(\rho-p).
\end{equation}
Within the gravity--thermodynamics conjecture, applying the first law of thermodynamics to the apparent horizon yields
\begin{equation}\label{FL}
    dE = T_h dS_h + WdV.
\end{equation}
The total energy enclosed by the apparent horizon is $E=\rho V$, with $V=(4\pi/3)\tilde{r}_{A}^{3}$ being the horizon volume, while $T_h$ and $S_h$ denote the apparent-horizon temperature and entropy, respectively. A direct differentiation yields
\begin{equation}\label{dE1}
 dE=4\pi\tilde
 {r}_{A}^{2}\rho d\tilde {r}_{A}+\frac{4\pi}{3}\tilde{r}_{A}^{3}\dot{\rho} dt.
\end{equation}
Using the conservation equation (\ref{Cons}), one finds
\begin{equation}
\label{dE2}
 dE=4\pi\tilde
 {r}_{A}^{2}\rho d\tilde {r}_{A}-4\pi H \tilde{r}_{A}^{3}(\rho+p) dt.
\end{equation}
We assume the apparent-horizon entropy is given by the generalized mass-to-horizon entropy \cite{Gohar2024},
\begin{equation}\label{Ent1}
S_{h}=\frac{2\pi n \gamma}{G(n+1)}\tilde {r}_{A}^{n+1}.
\end{equation}
Taking the differential of Eq.~(\ref{Ent1}) gives
\begin{eqnarray} \label{dSh}
dS_h&=& \frac{2\pi n \gamma}{G}\tilde {r}_{A}^{n} d\tilde {r}_{A}.
\end{eqnarray}
Substituting Eqs.~(\ref{T}), (\ref{Work2}), (\ref{dE2}), and (\ref{dSh}) into the first law of thermodynamics, Eq.~(\ref{FL}), yields the differential modified Friedmann equation,
\begin{equation} \label{Fried1}
n \gamma \tilde {r}_{A}^{n-4} d\tilde {r}_{A}= 4 \pi G H(\rho+p)
dt.
\end{equation}
Using the continuity equation (\ref{Cons}), we obtain
\begin{equation} \label{Fried2}
-2 n\gamma \tilde {r}_{A}^{n-4} d\tilde {r}_{A}= \frac{8 \pi G}{3}
d \rho.
\end{equation}
Integrating Eq.~(\ref{Fried2}) yields
\begin{equation} \label{Fried3}
\frac{2 n\gamma}{3-n} \tilde {r}_{A}^{n-3}  =  \frac{8 \pi
G}{3}\left(\rho+\rho_{\Lambda}\right),
\end{equation}
where $\Lambda$ is an integration constant identified with the cosmological constant, with $\rho_{\Lambda}=\Lambda/(8\pi G)$. Substituting $\tilde {r}_A$ from Eq.~(\ref{radius}) gives
\begin{equation} \label{Fried4}
\left(H^2+\frac{k}{a^2}\right)^{(3-n)/2} = \frac{4\pi G}{3 n
\gamma}(3-n) (\rho+\rho_{\Lambda}).
\end{equation}
Defining an effective gravitational constant as
\begin{equation} \label{Gef}
G_{\rm eff}=\frac{(3-n) G}{2 n \gamma},
\end{equation}
one can rewrite the modified Friedmann equation in the following form
\begin{equation} \label{Fried5}
\left(H^2+\frac{k}{a^2}\right)^{(3-n)/2} = \frac{8\pi G_{\rm
eff}}{3} (\rho+\rho_{\Lambda}).
\end{equation}
Throughout this work, we assume $\gamma>0$ and restrict the
entropy parameter to $0<n<3$, so that $G_{\rm eff}$ is positive
and finite. In the limiting case where
$n=\gamma=1$, one finds $G_{\rm eff}\rightarrow G$ and the
Friedmann equation (\ref{Fried5}) restores the result of standard
cosmology, as expected. Thus in comparison to the standard
cosmology here, we have two new parameters $n$ and $\gamma$. These
parameters can be constrained using cosmological observational
data. On the other hand for $n=\Delta+1$, the Friedmann equation
in Barrow cosmology is restored \cite{SheB1,SheB2}, while for
$n=2\delta-1$, it reduces to the modified Friedmann equation in
Tsallis cosmology \cite{SheT1}. The second modified Friedmann
equation can be easily derived by combining the first modified
Friedmann equation (\ref{Fried5}) with the continuity equation
(\ref{Cons}).
\section{The Model \label{cosm}}
This section constructs an HDE model within the modified cosmological framework derived from the generalized mass-to-horizon entropy relation. We restrict our analysis to a spatially flat universe ($k=0$); the extension to non-flat geometries is straightforward. For a flat FRW universe containing dynamical HDE, the modified Friedmann equation (\ref{Fried5}) takes the form
\begin{equation}\label{Fried6}
H^{3-n} = \frac{8\pi G_{\rm eff}}{3}\,\rho_{\rm tot},
\end{equation}
where \(\rho_{\rm tot} = \rho_m + \rho_D\) is the total energy
density (matter plus HDE). For \(n=\gamma=1\), the standard Friedmann equation is recovered. We set the integration constant
$\rho_\Lambda$ in Eq.~(\ref{Fried5}) to zero, so that the
dark-energy sector is described entirely by the HDE component.

The generalized mass-to-horizon relation \(M = \gamma
\frac{c^2}{G} L^n\) (with \(L\) an infrared cutoff) implies that
the energy contained within a sphere of radius \(L\) scales as \(M
\propto L^n\). The corresponding energy density is given by
\begin{equation}\label{rho1}
\rho_D = \frac{M}{\frac{4\pi}{3} L^3} \propto L^{n-3}.
\end{equation}
For convenience, we take the constant of proportionality to be
\begin{equation}\label{rho2}
\rho_D = \frac{3 c^2}{8\pi G_{\rm eff}} L^{n-3},
\end{equation}
where $c^2$ is a positive dimensionless constant. Clearly, for
$n=\gamma=1$, it restores the standard HDE
\cite{pav1,wang1,wang2}.
\subsection{Hubble Cutoff: Non-interacting case}
 We first identify the IR cutoff with the Hubble horizon, \(L =
H^{-1}\), which is a natural choice given its relation to the
apparent horizon in cosmology. This yields
\begin{equation}\label{LH}
\rho_D = \frac{3 c^2}{8\pi G_{\rm eff}} H^{3-n}.
\end{equation}
In the non-interacting scenario, dark matter and dark energy evolve independently. Matter (dark matter and baryons) is modeled as pressureless dust with energy density $\rho_m$ satisfying the standard continuity equations
\begin{eqnarray}
&&\dot{\rho}_m+3H\rho_m=0, \label{con1}\\ &&
\dot{\rho}_D+3H\rho_D(1+w_D)=0.\label{con2}
\end{eqnarray}
Dark energy is characterized by the equation of state parameter $w_D = p_D/\rho_D$. The dimensionless fractional density parameters are defined as
\begin{align}\label{omg}
    \Omega_{m}
    =\dfrac{\rho_{m}}{\rho_{cr}}=\dfrac{8\pi G_{\rm eff} \rho_{m}}{3H^{3-n}},\;\;\;\;\;\Omega_{D}
    =\dfrac{\rho_{D}}{\rho_{cr}}=\dfrac{8\pi G_{\rm eff}\rho_{D}}{3 H^{3-n}},
\end{align}
where $\rho_{cr}=3H^{3-n}/(8\pi G_{\rm eff})$ is the effective critical density. Consequently, the Friedmann equation (\ref{Fried6}) becomes
\begin{eqnarray}
 \Omega_{m}+ \Omega_{D}=1.
\end{eqnarray}
Following the approach of Ref.~\cite{Motaghi2024},
the background dynamics can be characterized by
$\Omega_D$, $w_D$, and $q$. For the Hubble-radius
cutoff, Eq.~(\ref{LH}) gives $\Omega_D=c^2$ and,
consequently, $\Omega_m=1-c^2$.

To obtain $w_D$ and the deceleration parameter $q =
-1-\dot{H}/H^2$, we differentiate $\rho_D$ with respect to time.
From $\rho_D \propto H^{3-n}$ it follows that
\begin{equation}
\frac{\dot{\rho}_D}{\rho_D} = (3-n)\frac{\dot{H}}{H}.
\label{eq:rhoDdot}
\end{equation}
Substituting this into the conservation law (\ref{con2}) yields
\begin{equation}
w_D = -1 - \frac{3-n}{3}\,\frac{\dot{H}}{H^2}. \label{wD1}
\end{equation}
Differentiating the Friedmann equation (\ref{Fried6}) with respect to cosmic time and inserting the continuity equations $\dot{\rho}_m = -3H\rho_m$ and $\dot{\rho}_D = -3H\rho_D(1+w_D)$ gives
\begin{equation}
(3-n)H^{2-n}\dot{H} = \frac{8\pi
G_{\rm eff}}{3}\bigl[-3H\rho_m -3H\rho_D(1+w_D)\bigr].
\end{equation}
Dividing by $3H^{3-n}$ and using the definitions of $\Omega_m$ and
$\Omega_D$ leads to
\begin{equation}
\frac{3-n}{3}\,\frac{\dot{H}}{H^2} = -\Omega_m - \Omega_D(1+w_D).
\label{eq:HdotH}
\end{equation}
Solving for $\dot{H}/H^2$ gives
\begin{equation}
\frac{\dot{H}}{H^2} = -\frac{3}{3-n}\bigl(1+c^2 w_D\bigr).
\label{Hdot1}
\end{equation}
Combining Eqs.~(\ref{Hdot1}) and (\ref{wD1}) yields $(1-c^2)w_D=0$. For a physically viable two-component cosmic fluid with positive energy densities, $0<c^2<1$, which strictly imposes $w_D=0$. Consequently, non-interacting HDE governed by the Hubble-radius cutoff behaves as pressureless dust. The deceleration parameter is then given by
\begin{equation}
q = -1 - \frac{\dot{H}}{H^2}
   = -1 + \frac{3}{3-n}. \label{q}
\end{equation}
In the standard limit ($n=1$), one recovers $q = 1/2$, corresponding to a decelerating matter-dominated regime. Because the positivity of $G_{\rm eff}$ requires $n<3$, it follows that $3/(3-n)>1$, which unconditionally yields $q>0$. Hence, the non-interacting model with $L=H^{-1}$ fails to generate cosmic acceleration, rendering a dark-sector interaction indispensable for driving late-time acceleration.
\subsection{Interacting case}
When an energy exchange between the dark sectors is allowed, the continuity equations are modified as~\cite{wang1}
\begin{eqnarray}
&&\dot{\rho}_m+3H\rho_m=Q, \label{consm}
\\&& \dot{\rho}_D+3H\rho_D(1+w_D)=-Q.\label{consq}
\end{eqnarray}
Following Ref.~\cite{Motaghi2024}, we adopt the phenomenological coupling $Q = 3b^2 H(1+r)\rho_D = 3b^2 H(\rho_m + \rho_D)$, with $b^2$ denoting the dimensionless transfer coefficient and $r \equiv \rho_m/\rho_D$ being the energy density ratio.

Within this framework, the holographic dark energy density and the critical density are defined respectively as
\begin{equation}
    \rho_D = \frac{3c^2}{8\pi G_{\rm eff}} H^{3-n}, \qquad \rho_{\rm cr} = \frac{3}{8\pi G_{\rm eff}} H^{3-n}.
\end{equation}
The fractional dark energy density is therefore given by
\begin{equation}
    \Omega_D \equiv \frac{\rho_D}{\rho_{\rm cr}} = c^2.
\end{equation}
Remarkably, $\Omega_D$ remains strictly constant throughout the cosmic evolution—even in the presence of interaction—owing to the identical $H^{3-n}$ scaling of both $\rho_D$ and $\rho_{\rm cr}$. Differentiating Eq.~(\ref{LH}) with respect to cosmic time and utilizing the continuity equation~(\ref{consq}) yields
\begin{equation}
\frac{\dot{H}}{H^2} = -\frac{3}{3-n} \left(1 + w_D + \frac{b^2}{c^2}\right),
\label{eq:Hdot}
\end{equation}
Combining this result with (\ref{Hdot1}) yields the equation of state parameter for the dark energy component
\begin{equation}
w_D = -\frac{b^2}{c^2(1-c^2)}.
\label{eq:wD}
\end{equation}
Thus, in the interacting case, $w_D$ is generally nonzero. As noted
above, $c^2$ is a positive, dimensionless constant and must satisfy
$c^2 \neq 1$ for the expression obtained here. When $b^2=0$, the
model does not yield accelerated expansion. Since $\Omega_D=c^2$,
we also have $d\Omega_D/dz=0$: the dark-energy density parameter
remains constant as the Universe expands. This behavior arises because, within this specific
model, both the dark energy density and the critical density share
an identical functional dependence on the Hubble parameter \(H\).

To obtain the deceleration parameter, we have
\begin{equation}
q = -1 + \frac{3}{3-n} \left( 1 - \frac{b^2}{1-c^2} \right)
\label{eq:q}
\end{equation}
For $b^2=0$, the non-interacting result in Eq.~(\ref{q}) is
recovered, while $b^2>0$ describes an interaction between the dark
sectors. The expansion is accelerated ($q<0$) when
$b^2>\frac{n}{3}(1-c^2)$ and decelerated ($q>0$) when
$b^2<\frac{n}{3}(1-c^2)$. The equality $b^2=\frac{n}{3}(1-c^2)$
gives $q=0$ and marks the transition point between the two phases.
\section{HDE with Granda-Oliveros as IR cutoff\label{GOasIR}}
In this section, we investigate the evolution of the cosmological
parameters of the mass-to-horizon model by adopting the
Granda--Oliveros (GO) cutoff as the infrared (IR) cutoff. The GO
cutoff has been previously employed in various HDE scenarios with
generalized entropies, including Tsallis HDE~\cite{Sharma2024,
Zadeh2018, TsallisInconsistency2026}, Barrow
HDE~\cite{Motaghi2024, Nandhida2022, Luciano2026},
Rényi~\cite{Sardar2025, Dixit2020, Prasanthi2020, RenyiFG2023,
Prasanthi2026, Mazumdar2026}, Kaniadakis~\cite{Cruz2026, Gonzalez2026,
Bouncing2025}, as well as the three-parameter (\(S_3\)) and
five-parameter (\(S_5\)) generalized entropies~\cite{Chen2026}.
The GO cutoff is defined as a local length scale that depends on
the Hubble parameter ($H$) and its time derivative ($\dot{H}$),
and is given by
\begin{equation}
L = \left( \alpha H^2 + \beta \dot{H} \right)^{-1/2},
\label{eq:GO_cutoff}
\end{equation}
where $\alpha$ and $\beta$ are dimensionless constants. This
prescription avoids causality issues and naturally incorporates
the dynamical evolution of the cosmic
expansion~\cite{Mahmoudifard2025,Luciano2026}. Thus, the dark
energy density is modified as follows
\begin{equation}
\rho_D = \frac{3}{8\pi G_{\rm eff}} \left( \alpha H^2 + \beta \dot{H} \right)^{\frac{1}{2}(3-n)}.
\label{eq:rho_D_GO}
\end{equation}
\subsection{Dark Energy Dominated Universe}
In this section, we consider the case in which dark energy
dominates the Universe. Consequently, the contributions of matter
and radiation to the field equations can be neglected. In this
regime, the Friedmann equation (\ref{Fried6}) takes the following
form
\begin{equation}
H^{3-n} = \frac{8\pi G_{\rm eff}}{3}\,\rho_D,
\end{equation}
Substituting the GO dark-energy density, Eq.~(\ref{eq:rho_D_GO}), into the Friedmann equation in the dark-energy-dominated limit, we obtain
\begin{equation}\label{Fried7}
H^{2} = \alpha H^{2} + \beta \dot{H},
\end{equation}
For $\beta\neq0$ and $\alpha\neq1$, the solution satisfying $H(t_0)=H_0$ is
    \begin{equation}\label{eq:H_general}
        H(t)=\frac{H_0}
        {1+\dfrac{\alpha-1}{\beta}H_0(t-t_0)}.
    \end{equation}
The corresponding scale factor is
    \begin{equation}
        a(t)=a_0\left[
        1+\frac{\alpha-1}{\beta}H_0(t-t_0)
        \right]^{\beta/(\alpha-1)},
    \end{equation}
on the branch where the expression in brackets is positive. For
$\alpha=1$, the expanding solution is de Sitter, with $H(t)=H_0>0$
and $a(t)=a_0e^{H_0(t-t_0)}$. For the subsequent analysis with
$\alpha\neq1$, it is convenient to shift the time origin. Defining
$t_*=t_0-\beta/[(\alpha-1)H_0]$ and relabeling $t-t_*$ as $t$,
Eq.~(\ref{eq:H_general}) takes the form
\begin{equation}\label{eq:H_solution}
    H(t)=\frac{\beta}{\alpha-1}\frac{1}{t}.
\end{equation}
Using $H=\dot a/a$, the scale factor can then be written as
$a(t)\propto |t|^{\beta/(\alpha-1)}$. The time domain is chosen to
ensure $H>0$; for $\beta>0$, this requires $t>0$ if $\alpha>1$ and
$t<0$ if $\alpha<1$. We use this shifted time coordinate in the
following.

To derive the equation-of-state parameter for MH cosmology with
the GO cutoff in the late-time, dark-energy-dominated Universe, we
differentiate Eq.~(\ref{eq:rho_D_GO}) and use the continuity
equation to obtain
\begin{equation}\label{eq:rho_dot}
\dot{\rho}_{D} =( \frac{3-n}{2} )\rho_{D} \frac{2\alpha H\dot{H}+\beta \ddot{H}}{\alpha H^{2}+\beta \dot{H}},
\end{equation}
Substituting the first and second time derivatives of $H(t)$
into Eq.~(\ref{eq:rho_dot}) yields
\begin{equation}\label{eq:rho_dot_final}
\dot{\rho}_{D} = \frac{n-3}{t} \rho_{D}
\end{equation}
Thus, the desired equation of state parameter is given by
\begin{equation}\label{eq:wde_final}
w_{D} = -1 + (\frac{3-n}{3}) \frac{\alpha - 1}{\beta},
\end{equation}
And the deceleration parameter for this case can be obtained as
\begin{equation}\label{eq:deceleration}
q = -1 - \frac{\dot{H}}{H^{2}} = -1 + \frac{\alpha - 1}{\beta}
\end{equation}
It is evident that $H$, $w_{D}$, and $q$ are determined solely by the parameters $\alpha$, $\beta$ and $n$.
\subsection{Non-interacting case}
We commence by assuming that the dark sectors of the cosmos, i.e.,
dark matter and dark energy, do not exchange energy and thus are
independently conserved. The modified Friedmann equation, along
with the continuity equations, retains the identical structure as
presented in Eqs.~(\ref{Fried6}), (\ref{con1}), and~(\ref{con2}),
albeit with the crucial distinction that the dark energy density,
$\rho_D$, now obeys Eq.~(\ref{eq:rho_D_GO})
\begin{eqnarray}
H^{3-n} = \frac{8\pi G_{\rm eff}}{3}(\rho_m &+& \rho_D),
\label{eq:53}
\\
\dot{\rho}_m + 3H\rho_m &=& 0,
\label{eq:continuity_m}
\\
\dot{\rho}_D + 3H\rho_D(1 + w_D) &=& 0.
\label{eq:continuity_D}
\end{eqnarray}
By dividing both sides of the modified Friedmann
equation~(\ref{eq:53}) by the revised dark energy density
$\rho_D$, and carrying out the necessary calculations, the
following expression is derived
\begin{equation}
\alpha + \beta \frac{\dot{H}}{H^2} =\Omega_D^{\frac{2}{3-n}}.
\label{eq:alpha_beta_relation}
\end{equation}

Taking the time derivative of both sides of the Friedmann
equation, and employing the continuity equations together with the
definitions $\Omega_D = \frac{8\pi G_{\rm eff} \rho_D}{3H^{3-n}}$
and $\Omega_m = \frac{8\pi G_{\rm eff} \rho_m}{3H^{3-n}}$, we
obtain
\begin{equation}
\frac{\dot{H}}{H^2} = -\frac{3}{3-n} \left(1 + \Omega_D w_D \right),
\label{eq:Hdot_H2}
\end{equation}
Substituting~(\ref{eq:Hdot_H2}) into~(\ref{eq:alpha_beta_relation}), we obtain
\begin{equation}
w_D = \frac{1}{\Omega_D} \left[ -\frac{3-n}{3\beta} \left( \Omega_D^{\frac{2}{3-n}} - \alpha \right) - 1 \right],
\label{eq:omega_D}
\end{equation}

The evolution of the fractional dark-energy density
can be expressed in terms of redshift using
\begin{equation}
\frac{d\Omega_D}{dz} = \frac{d\Omega_D}{dt} \cdot \frac{dt}{dz},
\label{eq:chain_rule}
\end{equation}
with
\begin{equation}
\begin{aligned}
\frac{d\Omega_D}{dt} &= \frac{d}{dt} \left( \frac{\rho_D}{\rho_{cr}} \right) \\
&= \frac{8\pi G_{\rm eff}}{3} \left( \frac{\dot{\rho}_D H^{3-n} - (3-n) H^{2-n} \rho_D \dot{H}}{H^{2(3-n)}} \right).
\end{aligned}
\label{eq:Omega_D_time_derivative}
\end{equation}
Combining Eq.~(\ref{eq:Omega_D_time_derivative})
with the continuity equation and $dt/dz=-1/[H(1+z)]$ yields
\begin{equation}
\frac{d\Omega_D}{dz} = \frac{\Omega_D}{1+z} \left[ 3(1+w_D) + \frac{3-n}{\beta} \left( -\alpha + \Omega_D^{\frac{2}{3-n}} \right) \right],
\label{eq:Omega_evolution}
\end{equation}
which simplifies to
\begin{equation}
\frac{d\Omega_D}{dz} = \frac{3\Omega_D}{1+z} \, w_D \, (1-\Omega_D).
\label{eq:Omega_evolution_2}
\end{equation}
Substituting Eq.~(\ref{eq:omega_D}) into Eq.~(\ref{eq:Omega_evolution_2}) gives the explicit
evolution equation,
\begin{equation}
\frac{d\Omega_D}{dz} = \frac{3(1-\Omega_D)}{1+z} \left[ -\frac{3-n}{3\beta} \left( \Omega_D^{\frac{2}{3-n}} - \alpha \right) - 1 \right].
\label{eq:Omega_evolution_final}
\end{equation}

Consequently, the deceleration parameter for the present model, within the framework of the Granda--Oliveros (GO) cutoff, is obtained as follows
\begin{equation}
\begin{split}
q &= -1 - \frac{\dot{H}}{H^2} \\
&= -1 + \frac{3}{3-n} \left( 1 + \Omega_D w_D \right) \\
&= -1 - \frac{1}{\beta}( \Omega_D^{\frac{2}{3-n}} - \alpha).
\end{split}
\label{eq:q_GO_final}
\end{equation}
We now plot the evolution of $\Omega_{D}$, $w_{D}$, and $q$ as
functions of redshift for the mass-to-horizon generalized entropy
with the GO cutoff for different values of the model parameters.
It should first be noted that, according to~\cite{Manoharan2024},
the values of $\alpha$ and $\beta$ lie in the ranges $[0.7, 1.0]$
and $[0.3, 0.9]$, respectively. Specifically, in this study we
adopt the observational results from four different sources for
the values of the model parameters. We take $\alpha=0.93$ and
$\beta=0.55$ from~\cite{Motaghi2024}, $\alpha=0.99$ and
$\beta=0.59$ from~\cite{Yarahmadi2024}, $\alpha=1.0$ and
$\beta=0.69$ from~\cite{Oliveros2022}, and $\alpha=1.03$ and
$\beta=0.56$ from~\cite{Luciano2026}. For the parameter $n$, we
use~\cite{Luciano2025}.

\begin{figure*}[t]
    \includegraphics[width=0.9\textwidth]{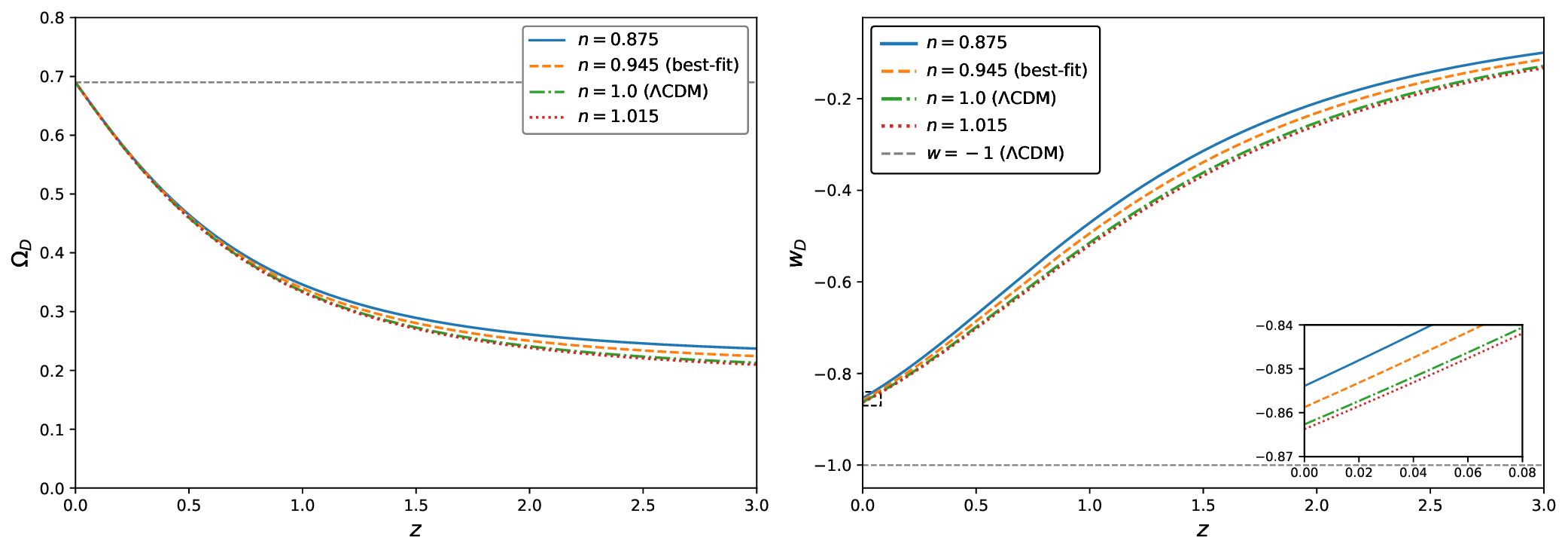}
    \caption{Evolution of the fractional energy density $\Omega_{D}$ (the left panel) and the DE-EoS parameter, $w_{D}$ (the right panel) as a function of redshift, $z$, for non-interacting HDE in a flat Mass-to-Horizon cosmology. Here, we set $\Omega_{D,0} = 0.69$, $\alpha = 1.03$ and $\beta = 0.56$.}
    \label{F1}
\end{figure*}

Fig.~\ref{F1} shows the dependence of the evolution of the DE
density parameter ($\Omega_{D}$) and the  equation-of-state
parameter ($w_{D}$) on the parameter $n$. As shown in the left
panel, the present-day value is $\Omega_{D,0}=0.69$ at $z=0$, from
which the curves decrease with increasing redshift and approach
$\Omega_{D}\simeq 0.24$ at $z=3$. This behavior is expected, since
matter dominated the universe at earlier times while the DE sector
was correspondingly smaller, and it has grown towards the present
epoch. For $0\leq z<0.6$, the curves are practically
indistinguishable, indicating a negligible dependence on the
parameter n. For $z>0.6$, however, they begin to diverge
gradually. The largest $\Omega_{D}$ is obtained for $n=0.875$ (the
blue curve), whereas increasing $n$ from $0.875$ to $1.015$
reduces $\Omega_{D}$. Moreover, the deviation of the MHR model
from the standard model ($n=1$) is more pronounced at earlier
epochs than at present.

\begin{figure*}[t]
    \includegraphics[width=0.55\textwidth]{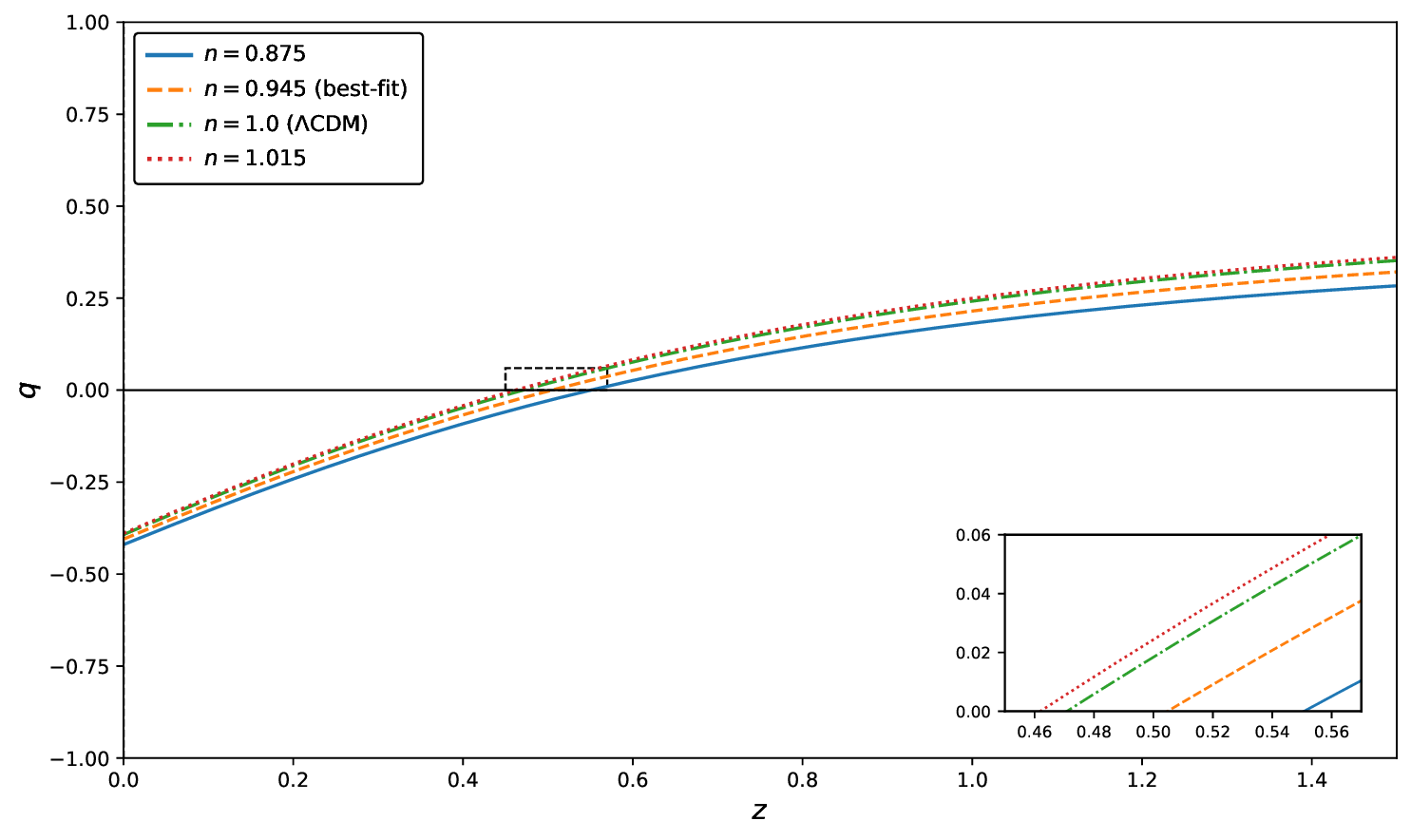}
    \caption{Evolution of the deceleration parameter $q$ against $z$, for a flat non-interacting Mass-to-Horizon cosmology. Here, we set $\Omega_{D,0} = 0.69$, $\alpha = 1.03$ and $\beta = 0.56$.}
    \label{F2}
\end{figure*}

\begin{figure*}[t]
    \includegraphics[width=0.9\textwidth]{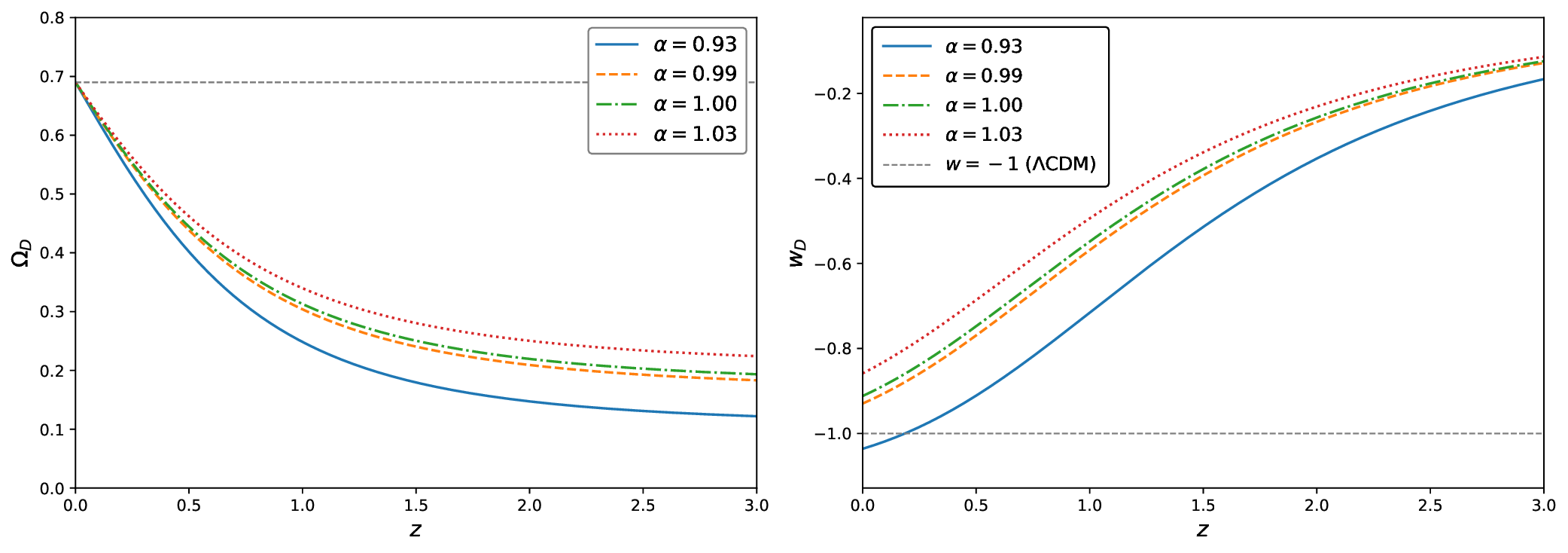}
    \caption{Evolution of $\Omega_{D}$ and the $w_{D}$ as a function of redshift, $z$, for non-interacting Mass-to-Horizon HDE in $\Omega_{D,0} = 0.69$, $n = 0.945$ and $\beta = 0.56$.}
    \label{F3}
\end{figure*}

\begin{figure*}[t]
    \includegraphics[width=0.9\textwidth]{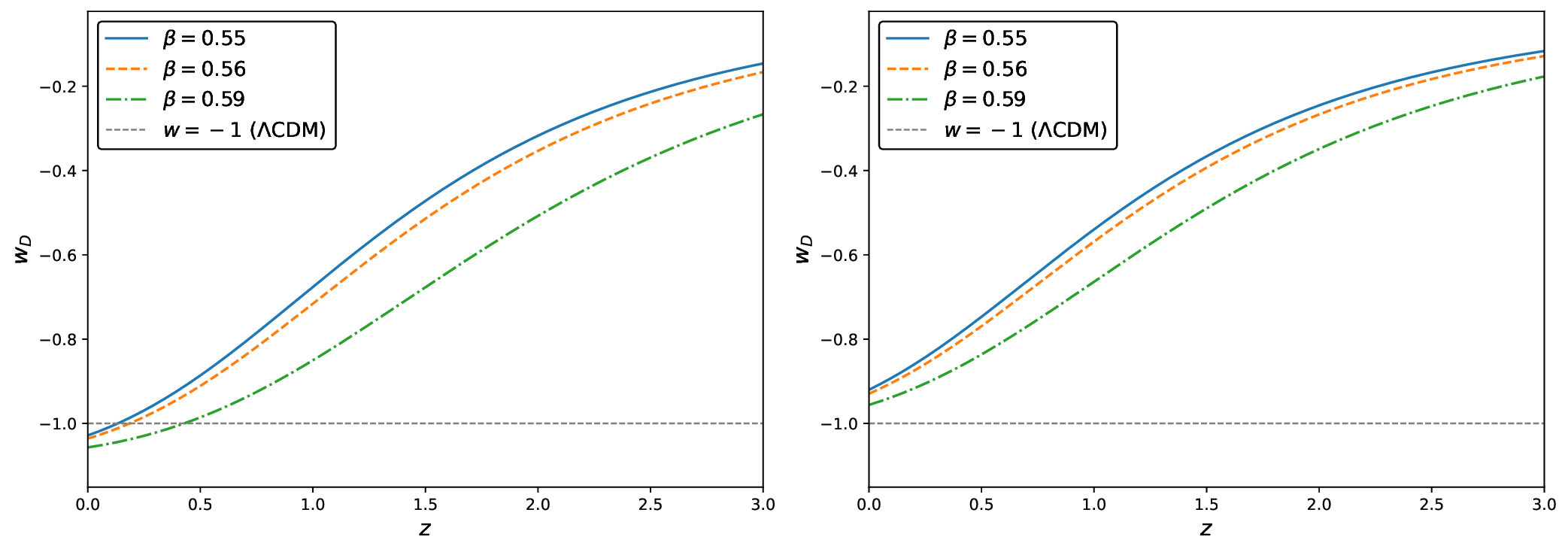}
    \caption{Evolution of $w_D$ as a function of $z$ for a non-interacting mass-to-horizon cosmology, with $\Omega_{D,0}=0.69$ and $n=0.945$. The left panel illustrates the evolution for $\alpha=0.93$, while the right panel corresponds to $\alpha=0.99$.}
    \label{F4}
\end{figure*}

\begin{figure*}[t]
    \includegraphics[width=0.9\textwidth]{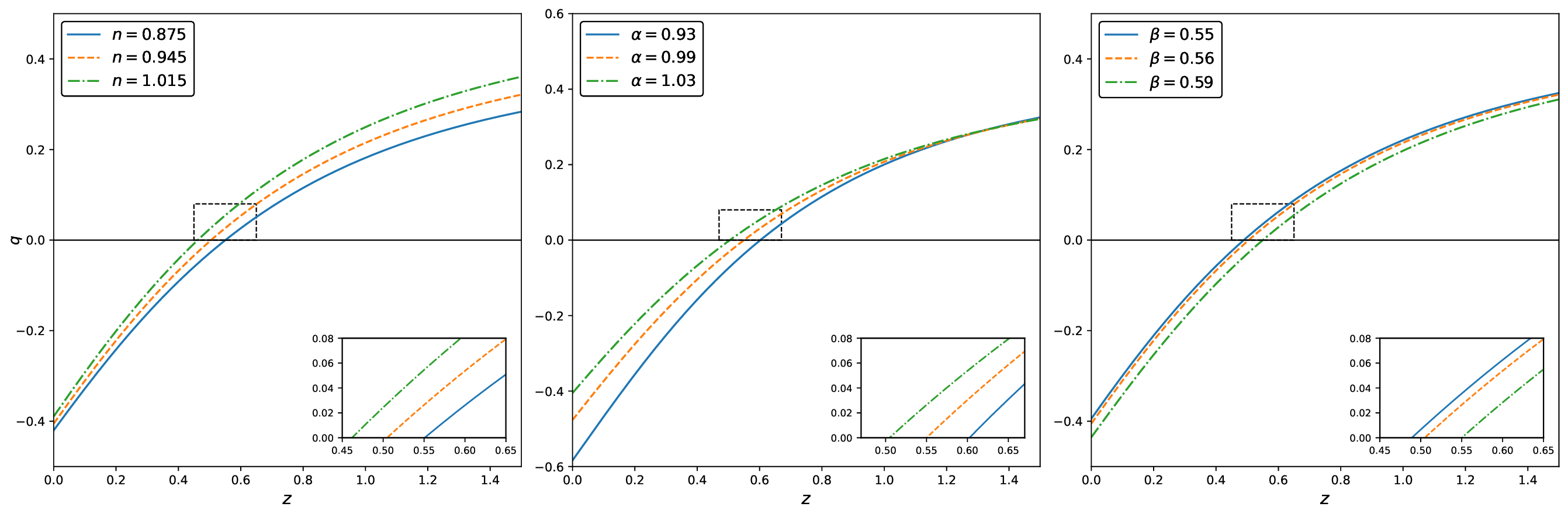}
    \caption{Evolution of $q$ as a function of $z$ for a non-interacting mass-to-horizon cosmology, with $\Omega_{D,0}=0.69$. The left panel varies $n$ (with $\alpha=1.03, \beta=0.56$); the middle panel varies $\alpha$ (with $n=0.945, \beta=0.56$); the right panel varies $\beta$ (with $n=0.945, \alpha=1.03$).}
    \label{F5}
\end{figure*}

The right panel shows that, at the present epoch ($z=0$), all
curves attain a value of approximately $w_D\simeq -0.86$,
indicating that DE in these models exhibits quintessence-like
behavior and deviates from the standard value $w_D=-1$. As one
moves towards the past (increasing $z$), the value of $w_D$
increases. This implies that DE behaved differently in the past
and departed from the standard model. The largest value of $w_D$
corresponds to $n=0.875$, whereas increasing $n$ leads to a
decrease in the EoS parameter. Figure \ref{F1} further shows that
varying the parameter $n$ has only a minor effect on the behavior
of the DE density, whereas its influence on the EoS parameter is
more pronounced. Fig.~\ref{F2} illustrates the effect of the
parameter $n$ on the deceleration parameter $q$. It is well known
that the universe is currently undergoing an accelerated
expansion. In this figure, $n=0.875$ (the blue curve) crosses the
transition point earlier than the others ($z=0.55$), while
$n=1.015$ crosses it later than the others ($z=0.46$). In other
words, the smaller the value of $n$, the earlier the phase
transition occurs.
Figures~\ref{F3}--\ref{F5} illustrate the sensitivity of the
background evolution to the GO parameters $\alpha$ and $\beta$.
Figure~\ref{F3} isolates the effect of $\alpha$ on $\Omega_D$ and
$w_D$, with $n=0.945$ and $\beta=0.56$ held fixed. As is evident,
the DE density for $\alpha=0.93$ (the blue curve) decreases with a
steeper slope. The smaller the value of $\alpha$, the more rapidly
DE has decreased in the past. In the right panel, the evolution of
$w_D$ is again examined for the same fixed values of $n$ and
$\beta$. For $\alpha=0.93$ (the blue curve), a transition occurs
in $w_D$, i.e., the curve crosses the phantom divide. Smaller
values of $\alpha$ correspond to lower values of $w_D$, with
$w_{D,0}\simeq-1.035$ for $\alpha=0.93$.

Figure~\ref{F4} compares the evolution of $w_D$ for different
values of $\beta$, with $\alpha=0.93$ in the left panel and
$\alpha=0.99$ in the right panel. This figure shows that $\alpha$
plays a more significant role in the evolution of the EoS
parameter $w_D$, such that for $\alpha=0.99$ at $z=0$ (the present
time) we have $w_D > -1$ in all cases, whereas for $\alpha=0.93$
we observe crossing of the phantom divide ($w_D < -1$) for all
values of $\beta$. In other words, whether $w_D$ is less than or
greater than $-1$ is strongly sensitive to $\alpha$. Moreover, in
both panels the difference between the sets of curves is
significant. The curve with $\beta=0.59$ moves towards zero at a
slower rate than the other two curves. Another conclusion that can
be drawn from this figure is that the parameter $\beta$ has a
stronger effect on the evolution of $w_D$ at earlier epochs
compared to the parameter $\alpha$: the smaller the value of
$\beta$, the more rapidly DE tends towards zero (matter-like
behavior).
Fig.~\ref{F5} shows the role of each of the model parameters ($n$,
$\alpha$, and $\beta$) in determining the time of the phase
transition from the decelerated to the accelerated phase. In the
left panel, $\alpha$ and $\beta$ are held fixed while $n$ is
varied ($\alpha=1.03$ and $\beta=0.56$). The smaller the value of
$n$, the larger the redshift at which the phase transition occurs
(closer to $z=0.6$), such that $n=1.015$ (the green curve) has the
latest phase transition (around $z=0.46$). In the middle panel,
the sensitivity of $q$ to variations of $\alpha$ is investigated
while $n$ and $\beta$ are held fixed ($n=0.945$ and $\beta=0.56$).
Increasing $\alpha$ shifts the transition from decelerated to
accelerated expansion to lower redshifts. For $\alpha=0.93$ (the
blue curve) the universe reaches the phase transition around
$z=0.6$. This value is consistent with the transition redshift
obtained in Ref.~\cite{Basilakos2025}, which is required by
observations and was derived within the GMHE framework. It is also
compatible with independent observational constraints on the
transition redshift, such as \(z_t = 0.64^{+0.12}_{-0.09}\) from a
joint Gaussian-process analysis of SN, BAO, and CC
data~\cite{Haridasu2018}, \(z_t = 0.61\) from a kinematic analysis
of the SNLS supernova sample~\cite{Cunha2008}, and \(z_t =
0.65^{+0.03}_{-0.19}\) obtained within the \(\Lambda(t)\)CDM
framework using CC+Pantheon data~\cite{Myrzakulov2025}. In the
right panel, the sensitivity of $q$ to $\beta$ is examined while
$n$ and $\alpha$ are held fixed ($n=0.945$ and $\alpha=1.03$). In
this figure, it can be seen that with increasing $\beta$, the
phase transition occurs at larger redshifts, and $\beta=0.59$ (the
green curve) reaches the phase transition earlier (around
$z=0.55$). It can also be concluded that, in all cases examined in
these three panels, the phase transition point lies in the
interval $[0.46, 0.60]$, which is in relative agreement with
current cosmological observations ($z_t=0.6$).
\subsection{Interacting case}

In the interacting case, differentiating the modified
Friedmann equation and using the continuity
equations~(\ref{consm}) and (\ref{consq}) yields
\begin{equation}
\frac{\dot{H}}{H^2} = -\frac{3}{(3-n)} \left(1 + \Omega_{D} \, w_{D}\right)
\label{eq:friedmann_interacting}
\end{equation}

\begin{figure*}[t]
    \includegraphics[width=0.9\textwidth]{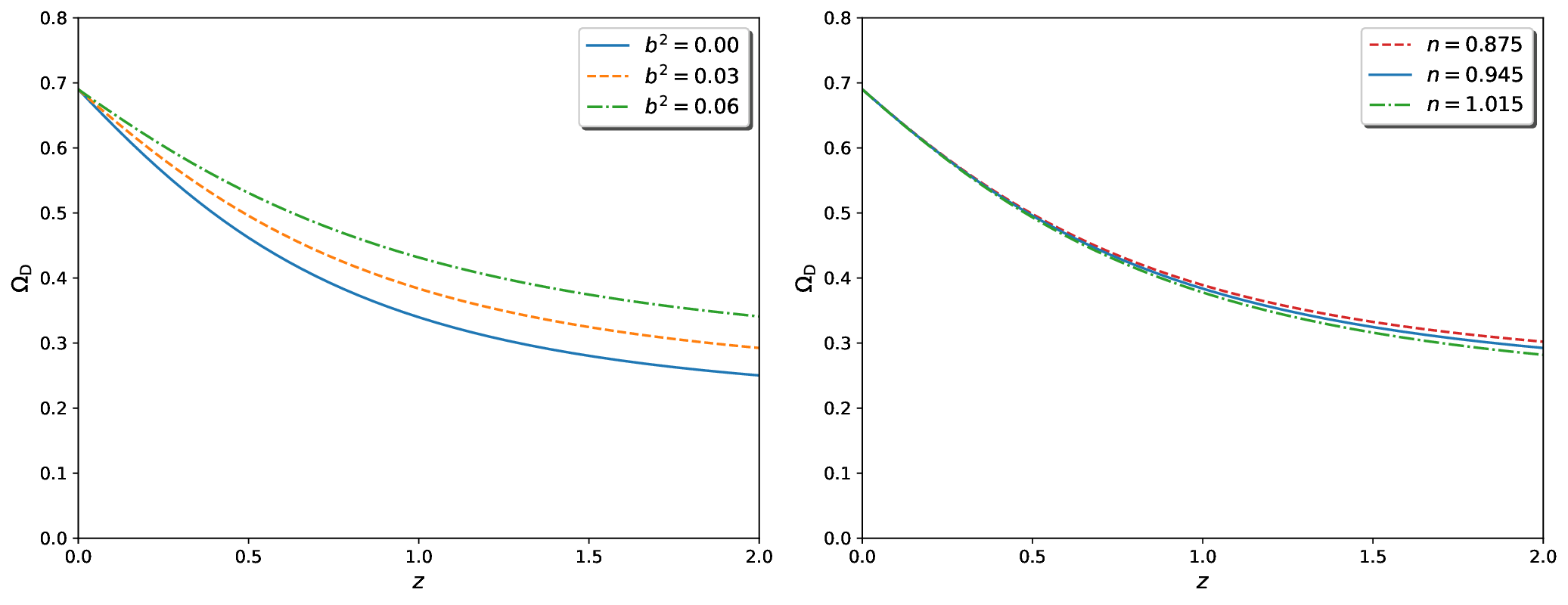}
    \caption{Evolution of $\Omega_{D}$ against $z$ for HDE for flat Interacting Mass-to-Horizon cosmology for different values of $b^{2}$ (left) and $n$ (right). Here, we have set $\Omega_{D,0}=0.69$, $\alpha = 1.03$, $\beta = 0.56$. In the left panel $n = 0.945$ and in the right one $b^{2} = 0.03$.}
    \label{F6}
\end{figure*}

In the next step, after dividing the Friedmann equation by $\rho_D$, the equation of state parameter takes the form
\begin{equation}
w_D = \frac{1}{\Omega_D} \left[ -\frac{(3-n)}{3\beta}  (\Omega_D^{\frac{2}{3-n}} - \alpha) - 1 \right]
\label{eq:eos_parameter}
\end{equation}
The corresponding evolution equation for the dark-energy density parameter is
\begin{multline}
\frac{d\Omega_D}{dz} = \frac{d\Omega_D}{dt} \frac{dt}{dz} \\
= \frac{3}{(1+z)} \Bigg[ (1-\Omega_D) \Bigg( -\frac{(3-n)}{3\beta} \Bigg( \Omega_D^{\frac{2}{3-n}} - \alpha \Bigg) - 1 \Bigg) + b^2 \Bigg]
\label{eq:evolution_Omega_interacting}
\end{multline}
The deceleration parameter for the interacting case, with the GO cutoff, is obtained as follows
\begin{equation}
q = -1 - \frac{\dot{H}}{H^2} = -1 - \frac{1}{\beta} \left(\Omega_D^{\frac{2}{3-n}} - \alpha \right)
\label{eq:deceleration_interacting}
\end{equation}
Figure~\ref{F6} shows the redshift evolution of $\Omega_D$ for
different values of $b^2$ and $n$. The left panel illustrates the
dependence on the interaction strength $b^2$. All curves coincide
at $z=0$ because the same initial condition, $\Omega_{D,0}=0.69$,
is imposed. As one moves towards the past, the DE density
decreases, but the rate of this decrease is strongly sensitive to
the value of $b^2$, and with decreasing $b^2$, $\Omega_D$
decreases. In the right panel, the effect of the parameter $n$ on
the evolution of $\Omega_D$ is investigated. The parameter $n$ had
a stronger influence on the evolution of the cosmic DE density in
the past, whereas at present this influence is not significant. It
should also be noted that the effect of the parameter $b^2$ on the
evolution of the DE density is more pronounced than that of $n$.

\begin{figure*}[t]
    \includegraphics[width=0.9\textwidth]{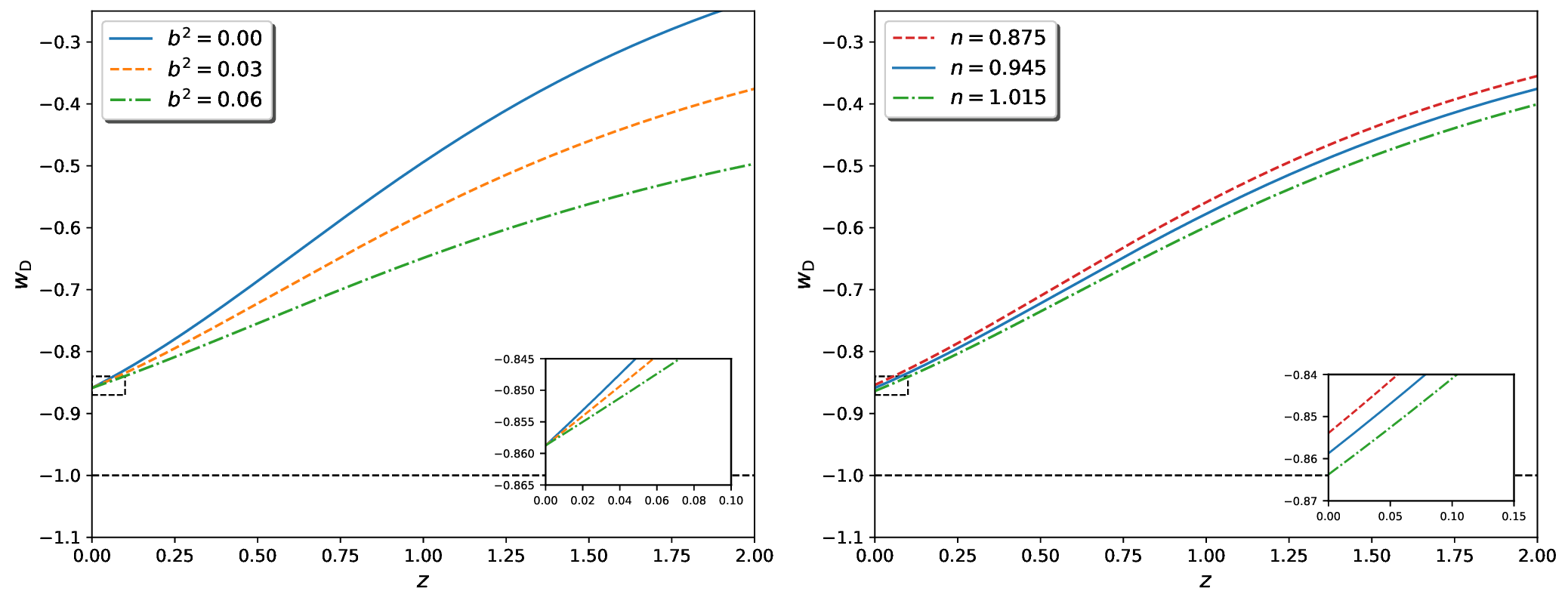}
    \caption{Evolution of $w_D$ as a function of $z$ for a spatially flat, interacting mass-to-horizon cosmology, shown for different values of $b^2$ (left panel) and $n$ (right panel). For this figure, we set $\Omega_{D,0}=0.69$, $\alpha=1.03$, and $\beta=0.56$. In the left panel, $n=0.945$, while in the right panel, $b^2=0.03$.}
    \label{F7}
\end{figure*}

Fig.~\ref{F7} compares the behavior of the $w_D$ versus $z$ for
different values of $b^2$ and $n$. What is noticeable is that all
curves at the present time lie between $-0.87$ and $-0.85$. This
implies that, at present, DE exhibits a Quintessence-like behavior
and deviates from the value $-1$. In the left panel, we see that
increasing $b^2$ causes DE to move more slowly towards matter-like
behavior ($w_D \to 0$) in the past. This behavior is consistent
with the $\Omega_D$ plot in the previous figure: models with
larger $b^2$ have a larger $\Omega_D$ in the past, and therefore
their $w_D$ changes more slowly. In the right panel, it can again
be observed that the effect of the parameter $n$ on $w_D$ is
weaker than that of $b^2$.

\begin{figure*}[t]
    \includegraphics[width=0.9\textwidth]{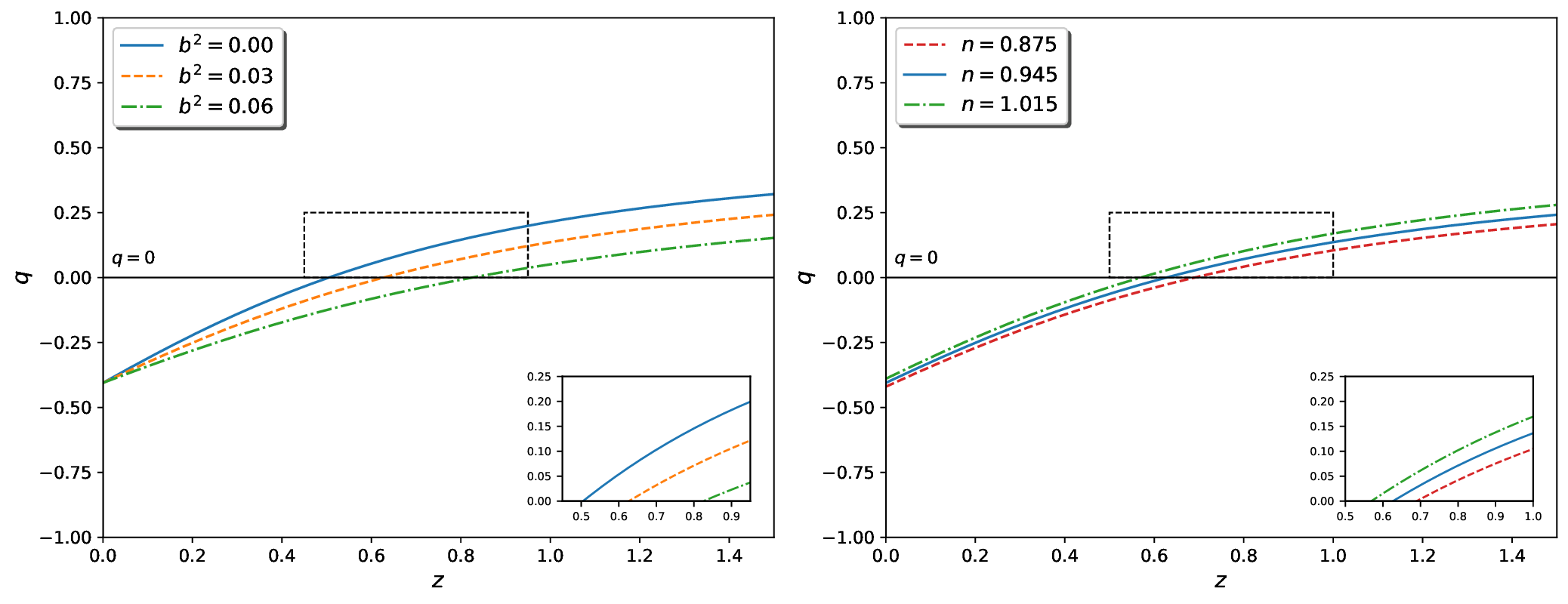}
    \caption{Evolution of  $q$ against $z$ for HDE in flat Interacting Mass-to-Horizon cosmology for different values of $b^{2}$ (left) and $n$ (right). Here, we have set $\Omega_{D,0}=0.69$, $n = 0.945$ (in left panel), $b^{2} = 0.03$ (in the right panel), $\alpha =1.03$ and $\beta = 0.56$.}
    \label{F8}
\end{figure*}

Fig.~\ref{F8} displays the effect of $b^2$ and $n$ on the
evolution of $q$. It can be observed that with increasing $b^2$,
the phase transition shifts towards larger redshifts: The case
$b^2=0.06$ has the earliest phase transition. As is evident from
the $\Omega_D$ plot, larger values of $b^2$ increase the
persistence of DE at higher redshifts. This greater persistence
means that DE can overcome dark matter earlier and drive the
universe towards the accelerated phase. Therefore, the phase
transition occurs at earlier epochs. Regarding the deceleration
parameter, we also see that the effect of $b^2$ on the phase
transition is far more pronounced than that of $n$.  It is worth
noting that both panels yield a transition redshift close to \(z
\approx 0.6\): in the left panel for \(b^2=0.03\), and in the
right panel for \(n=0.945\). This agrees with the transition
redshift discussed earlier in the text and with the
observationally favored transition redshift mentioned above.

\section{Stability and Statefinder of the model}\label{Stab}
\subsection{Stability Analysis}
\begin{figure*}[t]
    \includegraphics[width=1.0\textwidth]{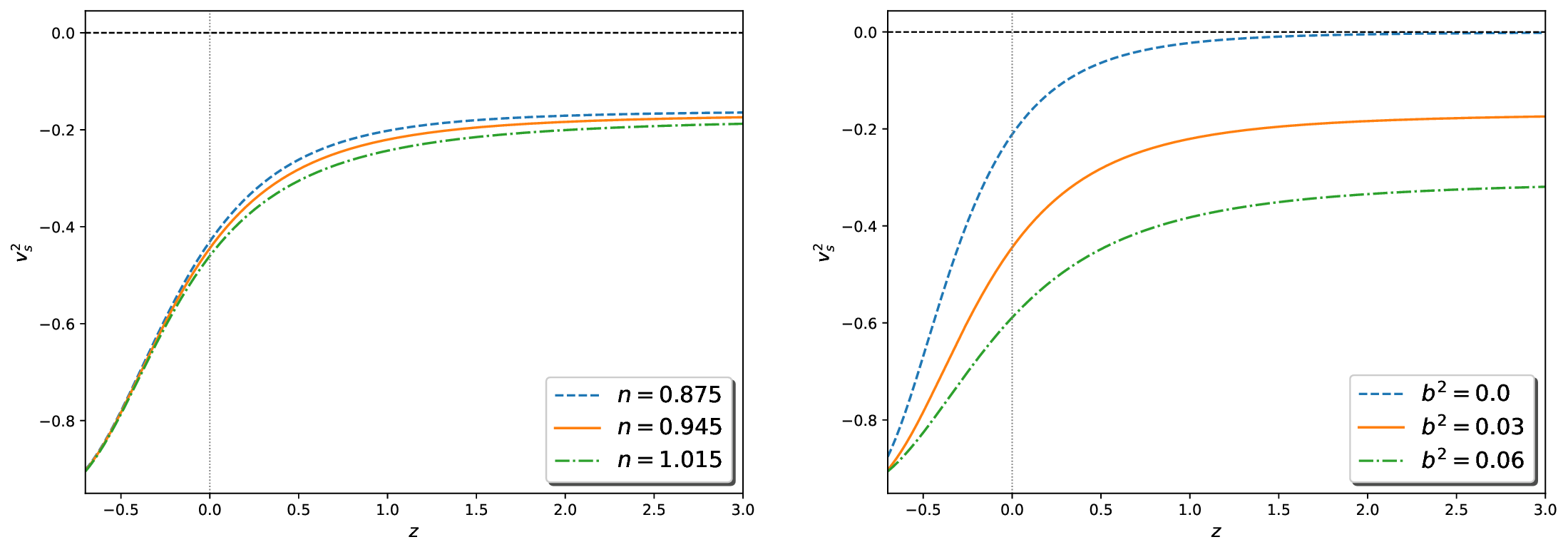}
    \caption{The squared sound speed $v_s^2$ versus redshift $z$ for the
        interacting HDE model in GMHE cosmology with the GO cutoff.
        Left panel: effect of the entropic parameter $n$, with fixed
        $\alpha=1.03$, $\beta=0.56$, $b^2=0.03$ and $\Omega_{D,0}=0.69$.
        Right panel: effect of the interaction coupling $b^2$, with
        fixed $n=0.945$, $\alpha=1.03$, $\beta=0.56$ and $\Omega_{D,0}=0.69$.
        The horizontal dashed line at $v_s^2=0$ separates the stable and
        unstable regions.}
    \label{F9}
\end{figure*}

\begin{figure*}[t]
    \includegraphics[width=1.0\textwidth]{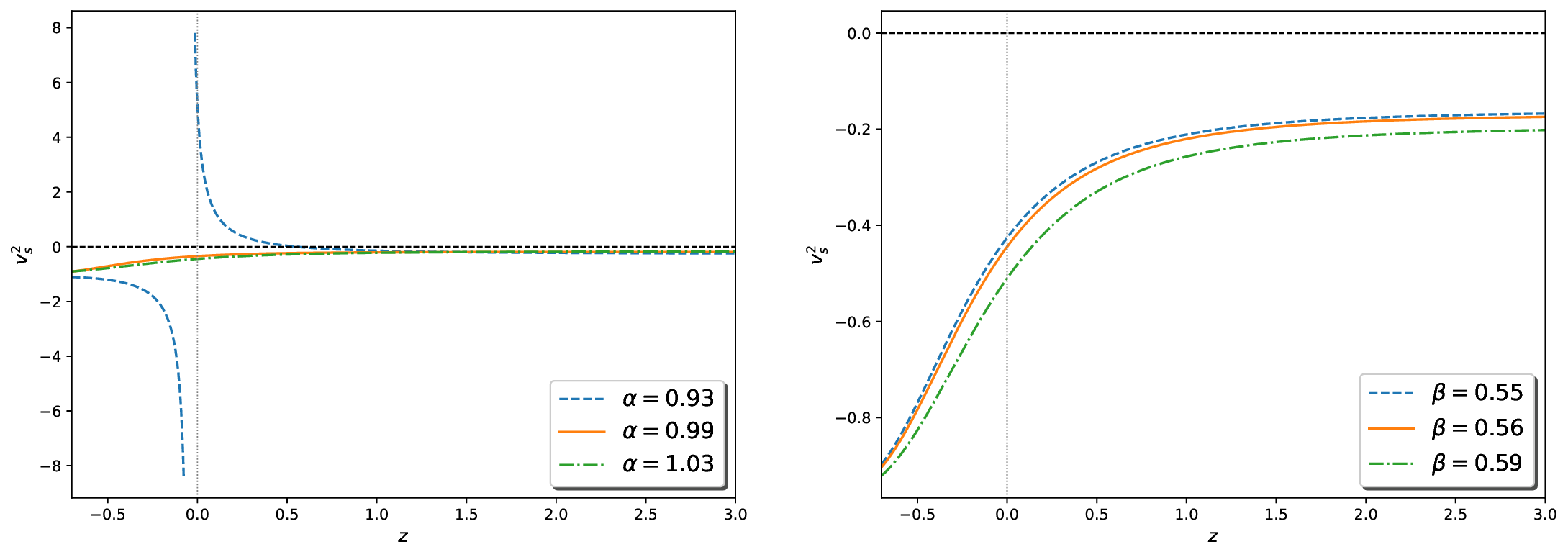}
    \caption{The squared sound speed $v_s^2$ versus redshift $z$ for the
interacting HDE model in GMHE cosmology with the GO cutoff.
Left panel: effect of the GO parameter $\alpha$, with fixed
$n=0.945$, $\beta=0.56$, $b^2=0.03$ and $\Omega_{D,0}=0.69$.
Right panel: effect of the GO parameter $\beta$, with fixed
$n=0.945$, $\alpha=1.03$, $b^2=0.03$ and $\Omega_{D,0}=0.69$.
The horizontal dashed line at $v_s^2=0$ separates the stable and
unstable regions.}
    \label{F10}
\end{figure*}

We assess the classical stability of the model using the
    squared sound speed $v_s^2$, a commonly used diagnostic in HDE
    models~\cite{Myung2007}. Positive $v_s^2$ indicates stability
    against small perturbations, whereas negative values signal a
    classical instability. From its definition,
\begin{equation}
v_s^2 = \frac{dp_D}{d\rho_D} = \frac{\dot{p}_D}{\dot{\rho}_D} = w_D + \frac{\rho_D}{\dot{\rho}_D}\,\dot{w}_D,
\label{eq:vs2}
\end{equation}
where \(p_D\) and \(\rho_D\) denote the pressure and energy density of the dark energy component, respectively, \(w_D = p_D/\rho_D\) is its equation-of-state parameter, and the dot represents differentiation with respect to cosmic time \(t\). Using $\dot{w}_D=(dw_D/d\Omega_D)\dot{\Omega}_D$ together with the interacting continuity equation and the evolution equation for $\Omega_D$,
Eq.~(\ref{eq:vs2}) becomes
\begin{equation}
v_s^2 = w_D + \frac{dw_D}{d\Omega_D}\,\frac{(1-\Omega_D)\left[-\frac{3-n}{3\beta}\left(\Omega_D^{\frac{2}{3-n}}-\alpha\right)-1\right]+b^2}{(1+w_D)+\frac{b^2}{\Omega_D}},
\label{eq:vs2-final}
\end{equation}
Figs.~\ref{F9} and \ref{F10} display $v_s^2(z)$ for different
values of the model parameters in the interacting case. In all
panels the dashed line $v_s^2=0$ separates the stable from the
unstable region.

Figure~\ref{F9} shows the effects of $n$ and $b^2$ on $v_s^2$. The left panel shows the dependence on the entropic parameter $n$. The three curves nearly
coincide: $v_s^2$ is negative over the entire redshift range and
approaches zero monotonically in the past, but never becomes
positive. The model is therefore classically unstable over the
entire redshift range, although it comes close to the stability
boundary at $z\gtrsim1$. Since $n$ controls the deviation of the
GMHE from the Bekenstein--Hawking area law and enters the
gravitational part of the Friedmann equation rather than the
dark-energy sector directly, its impact on $v_s^2$ is weak. Hence
$n$ is not a key parameter for classical stability.

The right panel of Fig.~\ref{F9} shows the effect of the
interaction coupling $b^2$. In contrast to $n$, this parameter has a
pronounced effect: increasing $b^2$ from $0$ to $0.06$ shifts $v_s^2$
towards more negative values and delays its approach to zero in the
past. Physically, a larger $b^2$ enhances the energy transfer between
the dark sectors, which amplifies dark-energy density fluctuations and
hence increases the classical instability. Thus $b^2$ is an important
stability parameter, although the model remains unstable at $z=0$ and in
the future for all values considered. It is also worth noting that the
case $b^2=0$ (non-interacting limit) yields the curve closest to the
stability boundary, confirming the destabilizing role of the interaction.

Figure~\ref{F10} shows the effects of $\alpha$ and $\beta$ on $v_s^2$. The left panel shows the dependence on the GO parameter $\alpha$. For $\alpha=0.99$ and $1.03$, $v_s^2$
stays smooth and close to zero over the whole redshift range,
indicating a marginally stable configuration. For $\alpha=0.93$,
however, a sharp feature appears near $z\approx0$ where $v_s^2$
temporarily becomes positive and large in magnitude. This occurs
because, for this combination of parameters, the denominator
$(1+w_D)+b^2/\Omega_D$ in Eq.~\eqref{eq:vs2-final} approaches
zero, so the linear response of pressure to density variations
becomes singular and the $v_s^2$ analysis breaks down. Among the
values considered, $\alpha=0.93$ therefore drives the model into a
regime where the standard linear stability analysis is no longer
applicable, whereas $\alpha=0.99$ and $1.03$ yield a well-behaved
model. A more detailed scan over a continuous range of $\alpha$
would be required to determine the precise boundary between the
stable and singular regimes; here we restrict our conclusions to
the discrete values examined.

The right panel of Fig.~\ref{F10} shows the effect of
$\beta$. Varying $\beta$ over the values $0.55$, $0.56$ and $0.59$ leaves
$v_s^2$ almost unchanged: the three curves nearly coincide, are negative
at $z\le0$, and approach zero from below in the past. Since $\beta$ is
the coefficient of the $\dot H$ term in the GO cutoff and mainly affects
the expansion rate rather than the perturbation dynamics, it is not a key
stability parameter, similar to $n$.

In summary, among the four parameters examined, $\alpha$ and
$b^2$ play a crucial role in the classical stability of the model, while
$n$ and $\beta$ have negligible effects. A detailed discussion of these
results and their implications is deferred to Section~VI.
\subsection{Statefinder Diagnostic}
We use the statefinder diagnostic to compare the model's
    expansion history with that of $\Lambda$CDM. Introduced by Sahni
    et al., this diagnostic involves the third time derivative of the
    scale factor $a(t)$~\cite{Sahni1}.
\begin{equation}
r = \frac{\dddot{a}}{a H^3},
\label{eq:statefinder-r}
\end{equation}
\begin{equation}
s = \frac{r-1}{3\left(q - \frac{1}{2}\right)},
\label{eq:statefinder-s}
\end{equation}
For the standard $\Lambda$CDM model, the Statefinder pair is given
by $\{r,s\} = \{1,0\}$. To assess the deviation of different dark
energy models from the standard model, one measures the departure
of the model's trajectory from the point $(1,0)$ in the $r$-$s$
plane. Expressing $r$ in terms of the deceleration parameter and
using $\dot{q}=(dq/d\Omega_D)\dot{\Omega}_D$ gives
\begin{equation}
r = 2q^2 + q - \frac{\dot{q}}{H} = 2q^2 + q + 3\frac{dq}{d\Omega_D} \Omega_D \left[ w_D (1-\Omega_D) + \frac{b^2}{\Omega_D} \right].
\label{eq:r-statefinder}
\end{equation}
Substituting relations~\eqref{eq:eos_parameter} and~\eqref{eq:deceleration_interacting} into the above expression, we obtain
\begin{equation}
\begin{split}
r ={}& 2\left(-1-\frac{1}{\beta}\left(\Omega_D^{\frac{2}{3-n}}-\alpha\right)\right)^2 \\
&+ \left(-1-\frac{1}{\beta}\left(\Omega_D^{\frac{2}{3-n}}-\alpha\right)\right) \\
&+ 3\left(-\frac{2}{3-n}\cdot\frac{1}{\beta}\,\Omega_D^{\frac{n-1}{3-n}}\right) \\
&\quad\times\Bigg[(1-\Omega_D)\bigg[-\frac{3-n}{3\beta}\left(\Omega_D^{\frac{2}{3-n}}-\alpha\right)-1\bigg]+b^2\Bigg],
\end{split}
\label{eq:r-statefinder-model}
\end{equation}
\begin{equation}
\begin{split}
s ={}& \frac{1}{-9\beta^2 - 6\beta\left(\Omega_D^{\frac{2}{3-n}}-\alpha\right)} \\
&\times\Bigg[4\left(\Omega_D^{\frac{4}{3-n}}-2\alpha\Omega_D^{\frac{2}{3-n}}+\alpha^2\right)
+ 6\beta\left(\Omega_D^{\frac{2}{3-n}}-\alpha\right) \\
&\quad - \frac{12\beta}{(3-n)}\,\Omega_D^{\frac{n-1}{3-n}} \\
&\quad \times\Bigg((1-\Omega_D)\bigg[-\frac{3-n}{3\beta}\left(\Omega_D^{\frac{2}{3-n}}-\alpha\right)-1\bigg]+b^2\Bigg)\Bigg],
\end{split}
\label{eq:s-general}
\end{equation}
\begin{figure*}[htbp]
    \includegraphics[width=0.9\textwidth]{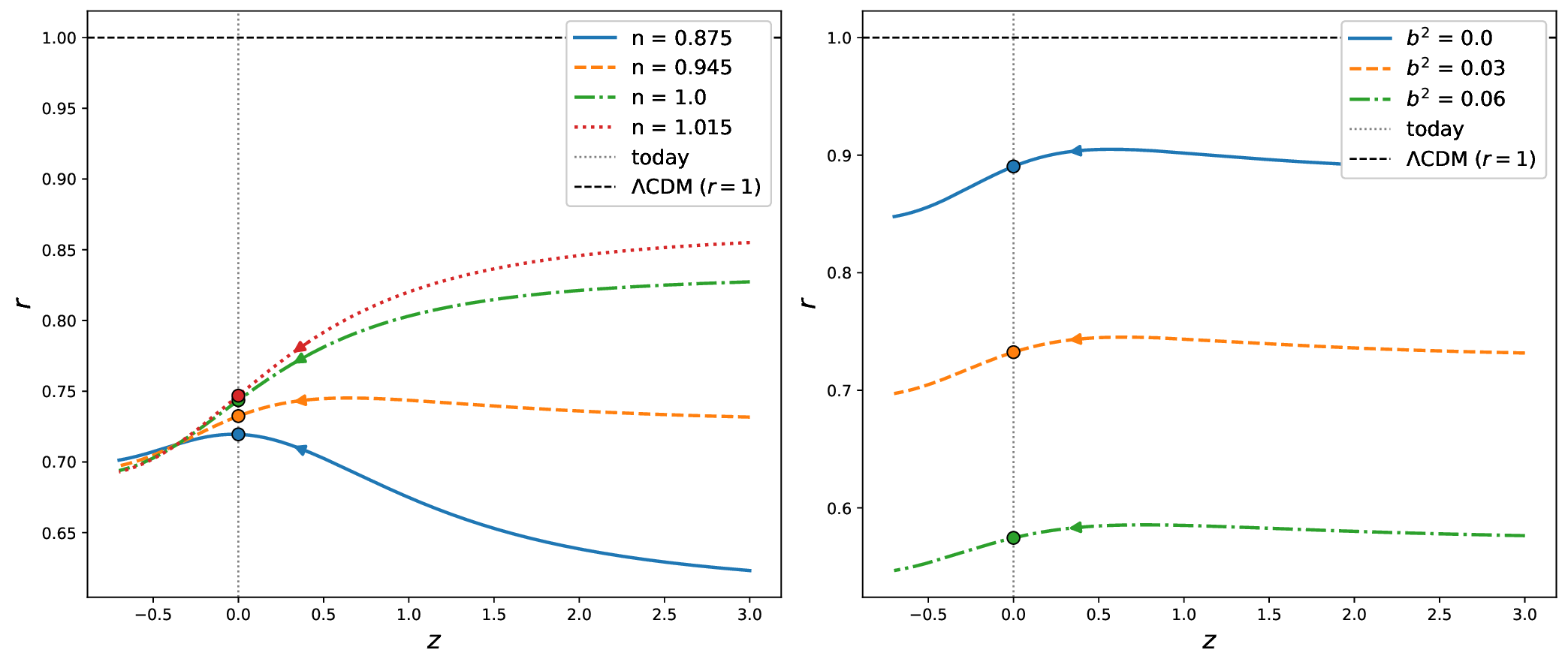}
    \caption{The evolution of $r$ against $z$ in the GMHE cosmology with
the GO cutoff for different values of $n$ (left panel) and $b^2$
(right panel). Here, we have set the fixed parameters
$\Omega_{D,0}=0.69$, $\alpha=1.03$ and $\beta=0.56$; for the left
panel $b^2=0.03$, and for the right panel $n=0.945$. The horizontal
dashed line at $r=1$ corresponds to the $\Lambda$CDM value.}
    \label{F11}
\end{figure*}
Eqs.~\eqref{eq:r-statefinder-model} and~\eqref{eq:s-general} show
the dependency of the statefinder pair \(\{r,s\}\) on the model
parameters of the GMHE model, namely \(\alpha\), \(\beta\), \(n\),
and \(b^2\). For \(n=1\), Eq.~\eqref{eq:eos_parameter} gives
\(\Omega_D = (\frac{2\alpha}{3\beta}-1)/(w_D +
\frac{2}{3\beta})\), while Eq.~\eqref{eq:deceleration_interacting}
reduces to \(q = -1 - \frac{1}{\beta}(\Omega_D - \alpha)\).
Substituting these expressions into Eq.~\eqref{eq:r-statefinder}
and taking the limits \(b^2 \to 0\) and \(w_D \to 0\), we find
\(q(w_D \to 0) = 1/2\), which immediately gives \(r=1\) and
\(s=0\). Therefore, the Statefinder parameters approach the
\(\Lambda\)CDM fixed point \(\{r,s\} = \{1,0\}\). On the other
hand, for \(n=1\) and \(b^2=0\), Eq.~\eqref{eq:r-statefinder}
simplifies to \(r = 1 + \frac{1-\alpha}{\beta}\left(3 +
\frac{2(\Omega_D-\alpha)}{\beta}\right)\), and
Eq.~\eqref{eq:r-statefinder-model} yields \(s =
2(\alpha-1)/(3\beta)\). Setting \(\alpha=1\) (with \(\beta \neq
0\)) again yields the \(\Lambda\)CDM fixed point \(\{r,s\} =
\{1,0\}\), which is consistent with the result obtained in
Ref.~\cite{Malekjani2011}.

Fig.~\ref{F11} shows the evolution of the statefinder parameter
$r$ against redshift $z$ for different values of $n$ (left panel)
and $b^2$ (right panel). At the present epoch, all trajectories
lie below the $\Lambda$CDM value ($r=1$). In the future, the
curves converge and slowly move away from $r=1$. Moving towards
the past, the behavior depends sensitively on $n$: the trajectory
for $n=0.875$ rapidly departs from $r=1$ and decreases
monotonically, while the curve for $n=0.945$ remains almost flat
and only slowly deviates from the $\Lambda$CDM value. The case
$n=1.015$ is the only one that approaches $r=1$ at higher
redshifts. (The curve for $n=1.0$ corresponds to the standard HDE
model and is shown for comparison.) The right panel reveals that
the interaction coupling $b^2$ has a pronounced effect: increasing
$b^2$ lowers the value of $r$ at all redshifts and drives the
trajectories further away from the $\Lambda$CDM fixed point.
Consequently, the non-interacting case ($b^2=0$) yields the
trajectory closest to $r=1$ over the entire redshift range.
\begin{figure*}[htbp]
    \includegraphics[width=0.9\textwidth]{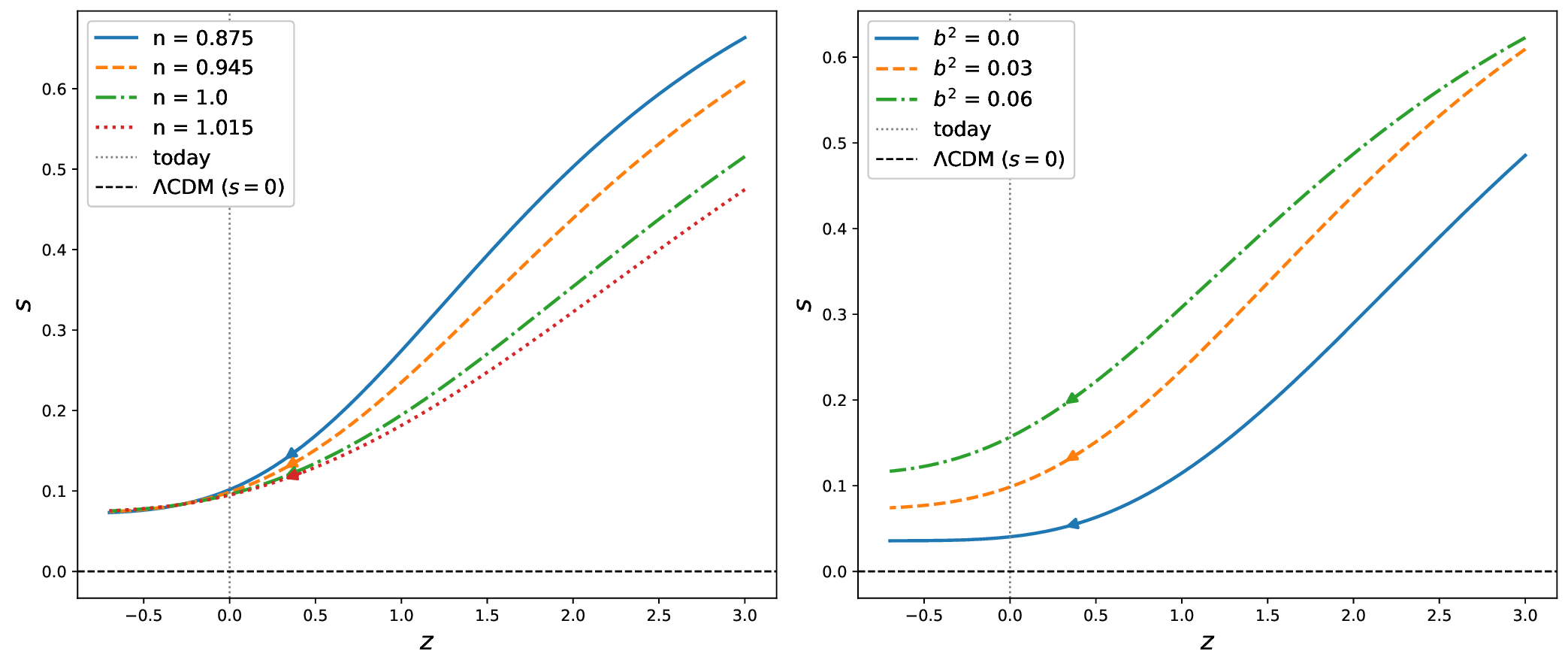}
    \caption{The evolution of $s$ against $z$ in the GMHE cosmology with
the GO cutoff for different values of $n$ (left panel) and $b^2$
(right panel). Here, we have set the fixed parameters
$\Omega_{D,0}=0.69$, $\alpha=1.03$ and $\beta=0.56$; for the left
panel $b^2=0.03$, and for the right panel $n=0.945$. The horizontal
dashed line at $s=0$ corresponds to the $\Lambda$CDM value.}
    \label{F12}
\end{figure*}

Fig.~\ref{F12} displays the evolution of the Statefinder parameter
$s$ against $z$. At the present epoch, $s$ is positive for all cases,
indicating a quintessence-like behavior. In the future, $s$ tends to
zero from above. Moving towards the past, $s$ increases monotonically
for all considered values of  $n$. In the left panel,
the model with $n=0.875$ exhibits the largest growth of $s$ at high redshifts, whereas increasing $n$
suppresses the value of $s$. The right panel demonstrates
that a larger interaction coupling $b^2$ results in a larger $s$ at all
redshifts, pushing the trajectories further into the region $s>0$.
\begin{figure*}[htbp]
    \includegraphics[width=0.9\textwidth]{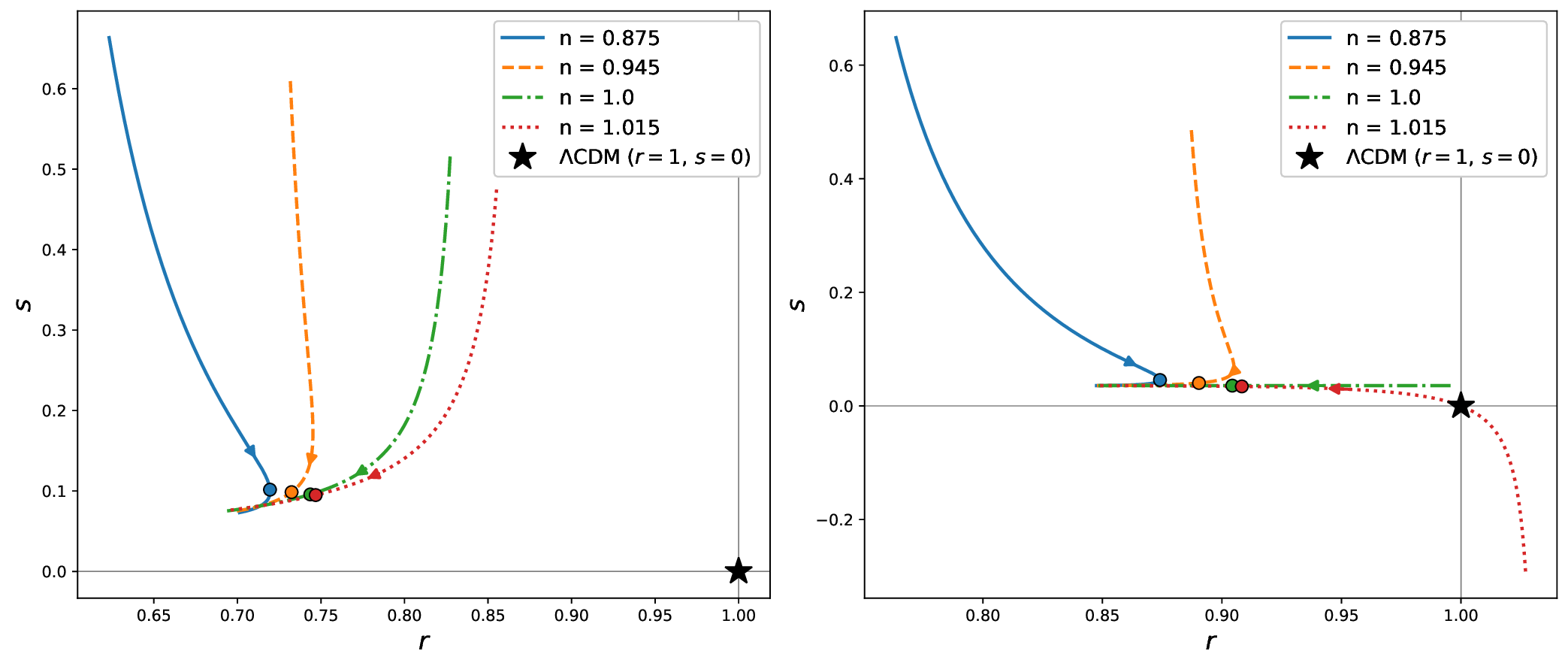}
     \caption{The evolution of $s$ against $r$ in the GMHE cosmology with
the GO cutoff for different values of $n$. The left panel corresponds
to $b^2=0.03$, while the right panel corresponds to $b^2=0$. Here, we
set $\Omega_{D,0}=0.69$, $\alpha=1.03$ and $\beta=0.56$. The black
star marks the $\Lambda$CDM fixed point $(r,s)=(1,0)$.}
    \label{F13}
\end{figure*}

\begin{figure*}[htbp]
    \includegraphics[width=0.9\textwidth]{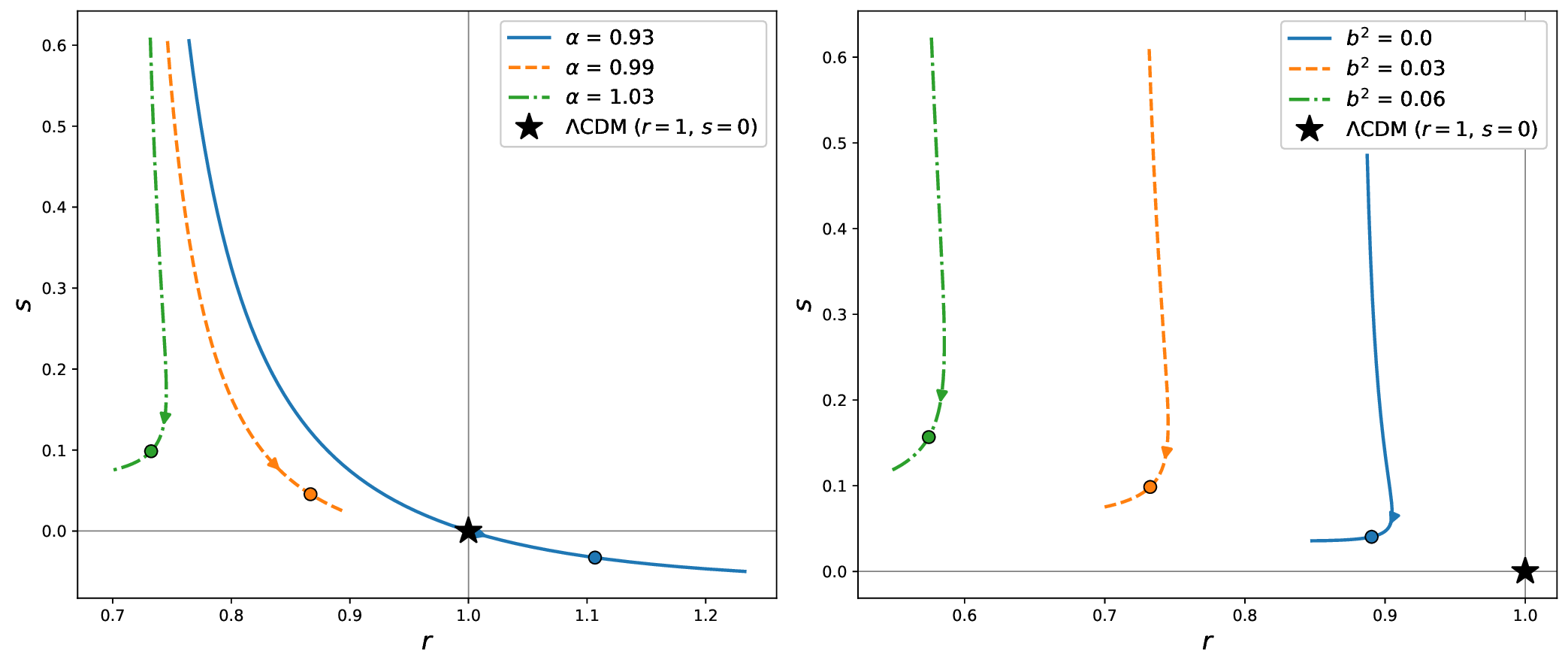}
    \caption{The evolution of $s$ against $r$ in the GMHE cosmology with
the GO cutoff for different values of $\alpha$ (left panel) and $b^2$
(right panel). In both panels we set $\Omega_{D,0}=0.69$, $n=0.945$
and $\beta=0.56$, with $b^2=0.03$ in the left panel and $\alpha=1.03$
in the right panel. The black star marks the $\Lambda$CDM fixed point
$(r,s)=(1,0)$.}
    \label{F14}
\end{figure*}
Fig.~\ref{F13} shows the effect of varying the parameter \(n\) on
the statefinder diagram for two values of \(b^2\) (left panel
\(b^2=0.03\) and right panel \(b^2=0\)). As shown in the figure,
decreasing \(b^2\) from \(0.03\) to zero brings the curves closer
to \(\Lambda\)CDM, indicating that the interaction drives the
model away from the standard cosmological behavior. The
trajectories in the \(r\)-\(s\) plane depend sensitively on the
parameter \(n\). In both panels, \(n=1.015\) exhibits the closest
behavior to \(\Lambda\)CDM (the case \(n=1.0\) corresponds to the
standard HDE model and is shown only for comparison). The parameter $n$ controls deviations from the
    Bekenstein--Hawking area law. In the parameter range considered,
    the larger deviation of $n$ from $1$ is accompanied by a greater
    departure of the statefinder trajectory from $(1,0)$. Therefore, \(n=0.875\) (with the
largest deviation from standard entropy) exhibits the farthest
behavior from \(\Lambda\)CDM. In the right panel, where the
interaction is switched off (\(b^2=0\)), the trajectory for
\(n=1.015\) passes exactly through the point \((1,0)\) in the
past. Before reaching this point, the model exhibits a
phantom-like behavior (\(s<0\)), and after crossing it, the
trajectory enters the quintessence region (\(s>0\)), where it
remains at the present time. Moreover, the interaction \(b^2\),
through energy transfer between dark matter and dark energy,
alters the dynamical evolution of the model and drives it further
from the standard path; this is why switching off the interaction
(\(b^2=0\)) brings the trajectories closer to \(\Lambda\)CDM.

Fig.~\ref{F14} further illustrates the roles of \(\alpha\) and
\(b^2\). The parameter \(\alpha\) in the Granda-Oliveros (GO)
cutoff plays a key role in determining the effective length scale.
In the left panel, the trajectory for \(\alpha=0.93\) (blue curve)
yields the closest behavior to \(\Lambda\)CDM at the present
epoch, and in the past, it passes exactly through the point
\((1,0)\). This means that, for this particular value, the
corrections from the GO cutoff and the GMHE entropy nearly balance
each other, and the model exhibits a nearly standard behavior.
Increasing \(\alpha\) to higher values breaks this balance and
causes the model to deviate from \(\Lambda\)CDM at the present
epoch. It is also important to note that for \(\alpha=1.03\), the
trajectory crosses the phantom divide (\(s=0\)) in the past,
entering the region \(s<0\), before returning to the quintessence
region at the present time. On the other hand, \(b^2\), as the
interaction coefficient, directly affects the energy exchange
between the dark sectors. Therefore, as seen in the right panel,
increasing \(b^2\) amplifies this exchange and consequently drives
the cosmic evolution further from the \(\Lambda\)CDM path.

In the \(r\)-\(s\) plane of Fig.~\ref{F14}, all the present-day
points (colored circles) lie in the region \(s > 0\), confirming
the quintessence-like nature of dark energy at the present epoch.
This is consistent with the findings of Ref.~\cite{Motaghi2024},
where the HDE model with the GO cutoff in Barrow cosmology also
exhibited a quintessence-like behavior in the past, crossing the
phantom divide at the present time. The quintessence-like behavior
(\(s > 0\), \(r < 1\)) suggests that the dark energy equation of
state in our model is bounded by \(w_D > -1\), which is consistent
with the latest observational constraints from Planck, DES, and
BAO data. Furthermore, although some trajectories (e.g.,
\(\alpha=1.03\) in Fig.~\ref{F14} and \(n=1.015\) with \(b^2=0\)
in Fig.~\ref{F13}) exhibit transient phantom crossings at higher
redshifts, none of them crosses the phantom divide (\(s = 0\)) at
the present epoch. This indicates that the model does not exhibit
stable phantom-like behavior at the present time, which is a
desirable feature from the perspective of theoretical consistency,
since phantom dark energy can lead to future singularities such as
the big rip. A comparison with the standard HDE model (with the
Hubble cutoff) is also instructive. In the standard HDE, the
statefinder trajectory typically deviates from \((1,0)\) more
strongly than in our model, because the standard HDE does not
include the entropy corrections encoded in the GMHE. Our model, by
contrast, can be tuned (via \(\alpha\)) to yield a trajectory very
close to \(\Lambda\)CDM at the present epoch, while still
exhibiting distinct behavior at higher redshifts. This makes the
GMHE-inspired HDE model a promising candidate for alleviating the
tension between early-time and late-time cosmological
observations, although a detailed observational analysis is
required to confirm this.

From an observational point of view, the statefinder diagnostic
provides a powerful tool to distinguish our model from
\(\Lambda\)CDM using forthcoming surveys such as Euclid, DESI, and
the Rubin Observatory. In particular, the deviation of the
trajectory from \((1,0)\) at \(z \sim 0.5\)--\(1\) could be probed
by these surveys, providing a direct test of the GMHE-inspired HDE
scenario. We leave a detailed observational analysis for future
work.
\section{Conclusions \label{Con}}
In this work, a holographic dark energy (HDE) model has been constructed within the modified cosmological framework dictated by the generalized mass-to-horizon entropy (GMHE). We examine how modifying the horizon entropy affects both the Friedmann equation and the holographic dark-energy density. The model was analyzed for two different infrared cutoffs, namely the Hubble radius and the Granda-Oliveros (GO) cutoff, in both non-interacting and interacting scenarios between dark matter and dark energy.

For the Hubble-radius cutoff, non-interacting HDE behaves as
pressureless matter (\(w_D = 0\)) and cannot reproduce the
late-time accelerated expansion, regardless of the value of the
model parameters. Introducing an interaction between dark matter
and dark energy, however, allows accelerated expansion when the
interaction strength exceeds the required threshold. In this case,
the dark energy density parameter \(\Omega_D\) remains constant,
and the deceleration parameter \(q\) depends on the interaction
coupling \(b^2\) and the parameter \(n\).

For the GO cutoff, the main focus of our analysis, three distinct
regimes were examined. In the dark-energy-dominated regime,
neglecting the contribution of matter and radiation, we solved the
Friedmann equation and obtained analytical expressions for the
Hubble parameter, the equation-of-state parameter, and the
deceleration parameter. These expressions show that the quantities
depend solely on the model parameters (\(\alpha\), \(\beta\), and
\(n\)) and are independent of the initial conditions, which
facilitates a straightforward comparison with observational data.
In the non-interacting regime, we analyzed the evolution of the
dark energy density parameter and the equation of state. For the
parameter choices considered, the present-day equation of state is
quintessence-like, with $w_D\approx-0.86$. At higher redshifts,
\(w_D\) increases, with the largest values corresponding to
smaller \(n\), while increasing \(n\) leads to a decrease in the
equation-of-state parameter. The density parameter decreases with
redshift, showing a negligible dependence on \(n\) at low
redshifts but a more pronounced divergence at higher redshifts.

The parameter $\alpha$ also has a significant influence on the
evolution of $w_D$; for smaller values (such as \(\alpha=0.93\)),
the model crosses the phantom divide at the present time, whereas
larger values keep \(w_D > -1\). The parameter \(\beta\) primarily
affects the evolution at earlier epochs. The model also exhibits a
transition from decelerated to accelerated expansion, with the
transition redshift approaching \(z \approx 0.6\) for appropriate
parameter choices (such as \(\alpha=0.93\)), which is consistent
with the value obtained from the observational data discussed
earlier in this work. In the interacting regime, the term
$Q=3b^2H(1+r)\rho_D$ modifies the evolution equation for the
dark-energy density parameter and significantly affects its
redshift dependence. At the present time, the equation of state
lies in the range between \(-0.87\) and \(-0.85\), confirming the
quintessence-like behavior. Increasing the interaction coupling
$b^2$ increases the dark energy density at higher redshifts and
shifts the equation of state more slowly toward matter-like
behavior. Notably, the effect of \(b^2\) on the evolution of both
the density parameter and the equation of state is more pronounced
than that of $n$ over the parameter ranges examined. In
particular, a stronger interaction causes dark energy to persist
at higher redshifts, leading to an earlier phase transition.
Specifically, for the interacting case with the GO cutoff, the
transition redshift can approach \(z \approx 0.6\) for suitable
parameter values (such as \(b^2=0.03\) and \(n=0.945\)), which is
in agreement with the observational constraints discussed earlier.

For the GO cutoff, the squared sound-speed analysis revealed a
stronger dependence on $\alpha$ and $b^2$ than on $n$ and $\beta$
over the parameter ranges examined. For \(\alpha = 0.93\), a sharp
singularity appears near \(z \approx 0\), originating from the
vanishing of the denominator in the expression for the squared
sound speed. This singularity signals the breakdown of the linear
stability analysis and calls for a nonlinear treatment of
perturbations. Switching off the interaction (\(b^2=0\)) brings
the curves closer to the stability boundary and shifts the
singularity to \(z \approx 0.25\), but does not eliminate the
fundamental instability associated with small \(\alpha\). This
result demonstrates the independence of the fundamental
instability from the interaction strength.

The statefinder analysis revealed that both the trajectories in
the $r$--$s$ plane and the redshift evolution of the individual
parameters are sensitive to $n$, $\alpha$, and $b^2$. Among the
values examined, $n=1.015$ yields the closest behavior to
$\Lambda$CDM, whereas $n=0.875$ shows the largest deviation,
confirming the role of \(n\) as the degree of deviation from the
standard entropy. This behavior is consistent with the evolution
of the individual Statefinder parameters, where the trajectory for
\(n=0.875\) departs most rapidly from the \(\Lambda\)CDM value,
whereas \(n=1.015\) remains closer to it at higher redshifts. In
the case \(b^2=0\), the trajectory for \(n=1.015\) exhibits a
phantom-like behavior (\(s<0\)) in the past, passes through the
point \((1,0)\), and enters the quintessence region (\(s>0\)) at
the present time, a feature visible in the corresponding \(s-z\)
diagram. Similar phantom-crossing behavior has been reported in
the literature for generalized entropic HDE
models~\cite{Chen2026}. The value \(\alpha=0.93\) provides the
closest match to \(\Lambda\)CDM at the present epoch and, in the
past, passes exactly through the point \((1,0)\). This indicates
that, for this particular value, a delicate balance is established
between the GO cutoff corrections and the GMHE entropy. Increasing
\(b^2\) drives the model further from \(\Lambda\)CDM, highlighting
the role of the interaction in deviating from the standard
behavior. This is reflected in the overall trend of the
Statefinder parameters, where a larger interaction strength
consistently pushes the trajectories away from the standard
cosmological path. Today, the trajectories lie in the region
\(s>0\) (quintessence-like behavior), and none of them crosses the
phantom divide (\(s=0\)) at the present epoch. This feature is
desirable from the perspective of theoretical consistency, as it
indicates that the model does not exhibit stable phantom behavior
currently and therefore avoids future singularities associated
with phantom dark energy. However, the model can exhibit transient
phantom behavior in the past without leading to a big rip (as
observed for \(n=1.015\) with \(b^2=0\)).

For suitable parameter choices, the statefinder trajectory
approaches the $\Lambda$CDM reference point at the present epoch
while retaining distinct behavior at higher redshifts. This may
indicate the potential of the model for alleviating the tension
between early-time and late-time cosmological observations. Our
findings are consistent with the results reported in
Ref.~\cite{Motaghi2024}, where a similar Statefinder analysis for
HDE with the GO cutoff in Barrow cosmology showed that increasing
the entropy parameter drives the model away from the
\(\Lambda\)CDM fixed point. This reinforces the idea that entropy
modifications, while offering a rich phenomenology, tend to push
the model away from the standard cosmological paradigm, and that a
delicate balance between the entropy parameters and the
interaction strength is required to achieve consistency with
observations.

Future work could constrain the model parameters using
cosmic-chronometer, baryon acoustic oscillation, and Type Ia
supernova data to assess the observational viability of
GMHE-inspired HDE. Furthermore, extending the analysis to non-flat
universes and to the perturbative level (e.g., growth of structure
and CMB anisotropies) would provide a more complete picture of the
model's observational signatures. Finally, exploring the
connection between the GMHE framework and other modified gravity
theories could offer new insights into the origin of cosmic
acceleration.
\acknowledgments{We thank Shiraz University Research Council. We
are also grateful to S. M. Movahed for reading the draft of this
paper.}

\end{document}